\documentclass[
 reprint,
 amsmath,amssymb,
 aps,
 longbibliography,
 prx
]{revtex4-2}

\usepackage{graphicx}    
\usepackage{dcolumn}     
\usepackage{bm}          
\usepackage{comment}
\usepackage{color,soul}

\usepackage{hyperref} 
\hypersetup{
    colorlinks=true,
    linkcolor=blue,
    filecolor=magenta,      
    urlcolor=cyan,
}

\usepackage{xcolor}

\definecolor{darkblue}{rgb}{0.0, 0.0, 0.55}       
\definecolor{darkturquoise}{rgb}{0.0, 0.81, 0.82} 
\definecolor{brown}{rgb}{0.65, 0.16, 0.16}

\begin{document}

\preprint{APS/123-QED}
\title{Nonlinear periodic orbit solutions and their bifurcation structure at the origin of soliton hopping in coupled microresonators}
\author{S. Deshmukh$^1$, A. Tusnin$^2$, A. Tikan$^{2,3}$, T. J. Kippenberg$^2$, T. M. Schneider$^1$}
\affiliation{%
$^1$Emergent Complexity in Physical Systems Laboratory (ECPS), Swiss Federal Institute of Technology Lausanne (EPFL), CH-1015 Lausanne, Switzerland\\
 $^2$Institute of Physics, Swiss Federal Institute of Technology Lausanne (EPFL), CH-1015 Lausanne, Switzerland\\
  $^3$Laboratoire Temps-Fréquence, Université de Neuchâtel, Avenue de Bellevaux 51, Neuchâtel, Switzerland \\
}%

\date{\today}

\begin{abstract}
Microresonator frequency combs, essential for future integrated optical systems, rely on dissipative Kerr solitons generated in a single microresonator to achieve coherent frequency comb generation. Recent advances in the nanofabrication of low-loss integrated nonlinear microresonators have paved the way for the exploration of coupled-resonator systems. These systems provide significant technological advantages, including higher conversion efficiency and the generation of dual dispersive waves. Beyond their practical benefits, coupled-resonator systems also reveal novel emergent nonlinear phenomena, such as soliton hopping—a dynamic process in which solitons periodically transfer between coupled resonators.

In this study, we employ a dynamical system approach and the corresponding well-established numerical techniques, extensively developed within the context of hydrodynamics and transitional turbulence, to investigate the bifurcation structure of periodic orbit solutions of the coupled Lugiato-Lefever equations that underlie soliton hopping in photonic dimers and trimers.
Our main finding uncovers a fundamental difference in the origin of the hopping process in dimers and trimers. We demonstrate that in dimers, hopping emerges from a branch of stable soliton solutions, whereas in trimers, it originates from an unstable branch. This distinction leads to a significant difference in pump power requirements, making the trimer system particularly promising for experimental investigation. 
We relate the bifurcation structure of the periodic orbits including their stability to the observed dynamics in simulated laser scans mimicking typical experimental investigations. Subcritical Hopf bifurcations of unstable equilibrium branches specifically explain observed hysteresis, the coexistence of multiple attractors at the same parameter values, and the importance of choosing a specific path in parameter space to reliably achieve a desired dynamical regime.
These findings provide crucial insights into the nonlinear dynamics of coupled multimode systems, with a particular focus on microresonators, offering a foundation for the controlled implementation of soliton-based technologies in future integrated photonic systems.

\end{abstract}

\maketitle

\section{Introduction}

Localized nonlinear structures play a crucial role across a variety of physical systems, ranging from the dynamics of surface water waves to light propagation in nonlinear media~\cite{Knobloch2015Spatial}. Under the appropriate approximations, such structures can be described by the same partial differential equation, the nonlinear Schr\"odinger equation (NLSE). 
The NLSE supports spatially localized solutions including solitons and rogue waves describing extreme wave events on the surface of oceans~\cite{elkoussaifi2018SpontaneousEmergenceRoguea}. 
As a Hamiltonian system, the NLSE is conservative and thus neither describes energy losses nor driving. 
Consequently, 
a driven-dissipative counterpart of the NLSE called the Lugiato-Lefever equation (LLE)~\cite{lugiato1987SpatialDissipativeStructures} has been widely investigated~\cite{Chembo2010Modal,Chembo2013Spatiotemporal,Godey2014Stability}. The LLE has specifically been shown to describe spatial patterns and their temporal dynamics in driven optical resonators including fiber-based Kerr microresonators~\cite{lugiato2018LugiatoLefeverEquation} as well as chip-integrated variants~\cite{kippenberg2018DissipativeKerrSolitons,pasquazi2018MicrocombsNovelGeneration}. Of special technological relevance are spatially localized structures, called ``dissipative solitons" (or Dissipative Kerr Solitons (DKS)~\cite{herr2014temporal} in the context of Kerr-nonlinear media). These nonlinear localized structures facilitate the generation of coherent, broadband and wide mode spacing optical frequency combs.
Such combs, creating optical spectra consisting of discrete regularly spaced spectral lines, form the basis of several optical high-precision measurement techniques as they allow precise mapping of optical to radio frequencies. Microresonator frequency combs (or microcombs) are thus foundational for the operation of many modern high-performance integrated photonic devices, and have enabled applications~\cite{sun2023ApplicationsOpticalMicrocombs} including on-chip photonic computing~\cite{Feldmann2021ParallelCore}, distance measurements~\cite{riemensberger2020massively,lukashchuk2021chaotic}, high-speed telecommunication~\cite{Marin-Palomo2017}, and many others~\cite{sun2023ApplicationsOpticalMicrocombs}.

While most integrated photonic devices had been based on single isolated microresonators, significant recent advancements in nano- and micro-fabrication techniques~\cite{liu2021high} now allow to manufacture arrangements of multiple resonators, whose enhanced quality factors, unprecedented levels of reproducibility and thus resonance frequency control, as well as reduced packing distance on a single chip, enable efficient coupling between the resonators. Those coupled optical resonator arrangements are increasingly gaining  attention~\cite{tikan2021emergent,tikan2022protected,helgason2021dissipative,ji2023EngineeredZerodispersionMicrocombsa,yuan2023SolitonPulsePairs} due to their potential in at least three areas. Firstly, coupled resonator systems may have performance advantages compared to their single-ring counterparts in the same application. Examples include enhanced comb generation efficiency~\cite{helgason2023SurpassingNonlinearConversion}, robust access to microcombs in normal dispersion cavities~\cite{xue2015NormaldispersionMicrocombsEnabled,helgason2021DissipativeSolitonsPhotonic}, protection from avoided mode crossings~\cite{tikan2022protected}, precise dispersion control for CMOS-ready Si$_3$N$_4$ photonics~\cite{ji2023EngineeredZerodispersionMicrocombsa,yuan2023SolitonPulsePairs} and voltage-controlled OPOs~\cite{pidgayko2023voltage}. 
Secondly, coupled resonators allow for applications that have no counterpart within a single resonator. An example are coupled resonator optical waveguides (CROWs) that enable applications ranging from optical filtering and slow light delay lines to nonlinear signal processing and light trapping~\cite{Ghosh2024,Morichetti2012First}. 
Thirdly, coupled resonator systems support novel dynamical phenomena. An example are periodic energy oscillations between two coupled microresonators, termed \emph{soliton hopping}~\cite{tikan2021emergent}. Such emergent nonlinear phenomena, not present in a single resonator, provide new opportunities to study pattern-formation mechanisms in dissipative nonlinear physical systems and may also enable the invention of novel photonic devices with yet-to-be envisioned future applications. Consequently, a thorough theoretical understanding of coupled optical resonators and their dynamics is required.    

Experimentally observed emergent spatiotemporal phenomena in coupled resonators correspond to nonlinear, self-organized patterns in the optical field, shaped by the interplay of dispersion, nonlinearity, coupling, driving, and dissipation. A central challenge is to understand the mechanisms by which such patterns emerge, how they depend on external control parameters such as pump power and detuning, and which dynamical regimes are accessible in different regions of parameter space. A systematic approach to address these questions is through the bifurcation analysis of fully nonlinear invariant solutions of the governing equations, which capture the stationary and time-periodic states underlying the observed patterns.

Computing exact invariant solutions and characterizing their bifurcation structure, has a long and successful history in the study of hydrodynamic pattern-forming systems~\cite{Hopf1948, Nagataf1990, Kerswell2005, Gibson2008, Kawahara2011, Suri2017, Graham2021, Reetz2019a, Crowley2022}. The study of invariant solutions forms the basis of fully nonlinear dynamical systems approaches and augments linear stability analysis as well as weakly nonlinear theory to explain the emergence and evolution of increasingly complex self-organized structures. 
A prototypical hydrodynamic pattern forming system is Rayleigh-Bénard convection~\cite{Bodenschatz2000, Lappa2009}, the flow of a Newtonian liquid confined between two parallel horizontal plates, heated from below and cooled from above. When the thermal driving characterized by the nondimensional Rayleigh number is increased, the unpatterned homogeneous conduction state loses stability to spatially periodic convection rolls via a primary bifurcation, akin to a modulation instability in optics. The corresponding instability onset can be captured by a linear stability analysis of the Oberbeck–Boussinesq equations, which determines the threshold parameter value and the dominant unstable modes responsible for the instability~\cite{Busse1978}. Close to the linear threshold, the resulting saturated patterns can often be described by weakly nonlinear theory yielding amplitude equations that capture the pattern amplitude. Linear stability analysis of the weakly nonlinear pattern solution can uncover secondary instabilities of the pattern that lead to the formation of more complex time-dependent or spatially modulated patterns~\cite{Busse1979, Subramanian2016}. 

However, farther from the instability threshold, in the strongly nonlinear and often weakly chaotic regimes, linear and weakly nonlinear arguments alone no longer provide a satisfactory explanation for many observed flow patterns. Instead, non-chaotic exact invariant solutions of the full 3D Navier-Stokes equations including equilibria, traveling waves, and periodic orbits have been shown to capture the dominant flow structures~\cite{Reetz2020e, Reetz2020, Zheng2024, Zheng2024b}.
Within the high dimensional state space of the system spanned by all instantaneous flowfields, these exact invariant solutions correspond to low-dimensional sets that are invariant under the time evolution. Invariant solutions can be dynamically stable or unstable. While the system displays turbulent dynamics, the state-space trajectories transiently approach the unstable invariant solutions embedded within the turbulent attractor along their stable manifolds and are redirected along their unstable ones, producing intermittent signatures of the solution patterns in space and time. Since these exact invariant solutions capture recognizable coherent flow structures, they are often also referred to as 'exact coherent states' or ECS. 
The collection of coexisting unstable solutions and their invariant manifolds constitutes a dynamical scaffold or backbone for the turbulent attractor, providing a global organizing framework for understanding pattern formation in the weakly turbulent regimes. Thereby, identifying invariant solutions and tracking their bifurcations with respect to key control parameters enables a global view of the system's dynamics—revealing the instability mechanisms by which patterns emerge or lose stability. We can specifically identify regions of multistability, where different competing attracting states coexist and result in hysteretic system behavior.

In the context of nonlinear optics,, various exact equilibrium solutions of the LLE on a periodic domain have been identified and characterized. These include Turing patterns~\cite{Qi2019,Parra-Rivas2018a} and spatially localized structures such as dissipative solitons~\cite{parra-rivas2018BifurcationStructureLocalized,Barashenkov1998Bifurcation,Cyril2017,Barashenkov1996Existence}. Of particular relevance is the bifurcation structure of the soliton solution branches, including homoclinic snaking and foliated snaking phenomena that describe sequences of saddle-node bifurcations that lead to the coexistence of localized states of different spatial extent~\cite{parra-rivas2018BifurcationStructureLocalized}.

For coupled LLEs, there have been relatively fewer developments based on the dynamical systems approach. For two coupled resonators, the soliton hopping dynamics has recently been studied within a mean-field description resulting from averaging the optical field in each resonator over space~\cite{Yelo-Sarrion2021Self-Pulsing,Yelo-Sarrion2022Self-Pulsing}. In this so-called driven dissipative Bose-Hubbard dimer self-sustained out-of-phase oscillations of the two mean-field amplitudes, resembling soliton hopping within the space-averaged description, emerge beyond a critical parameter threshold. For this system, described by coupled ODEs, a periodic orbit (or limit cycle) has been shown to bifurcate off the stationary state in a Hopf bifurcation~\cite{Yelo-Sarrion2022Self-Pulsing}. In a one-dimensional regular lattice of coupled resonators, generalized soliton hopping in the form of a soliton marching along the lattice has also been observed~\cite{Tusnin2023Nonlinear}. While soliton hopping appears to be a robust emergent phenomenon in coupled-driven optical resonators, a precise description in terms of time-periodic solutions of coupled LLEs describing the space- and time-dependent optical fields in all coupled resonators remains unexplored. 

In this article, we explore the spatiotemporal dynamics in coupled LLEs representing resonator dimers and trimers, using a dynamical systems approach. Augmenting direct numerical simulation, we specifically compute the stable and unstable exact invariant solutions, including equilibria and time-periodic solutions that underlie various spatio-temporal patterns in the dimer and trimer systems, continue their solution branches in parameter space, and elucidate their bifurcation structure.  
Both dimers and trimers display time-periodic behaviour supported by exact time-periodic solutions of the coupled nonlinear partial differential equations. However, when varying the detuning parameters mimicking typical experiments, the sequence of observed dynamical regimes at the onset of time-periodic soliton hopping differs in both systems. When the detuning frequency parameter is slowly increased and crosses the resonator resonance frequency for fixed pump power, dimer hopping is preceded by the observation of a stable supermode soliton. In contrast, trimer hopping emerges for increasing detuning parameter without prior observation of a stable soliton. Moreover, soliton hopping in trimers requires a considerably lower pump power than in dimers, suggesting that trimer hopping is easier to achieve in experiments.

To precisely identify the parameter regimes at which the dynamics can show soliton hopping, to understand which path in parameter space leads to robust temporal periodicity and to characterize the onset of time-periodic behavior in parameter sweeps, we perform a bifurcation analysis of the exact invariant solutions underlying both stationary supermode solitons and time-periodic soliton hopping.
Our analysis reveals that both dimer and trimer hopping are supported by branches of periodic orbits, that bifurcate off an equilibrium branch of stationary supermode solitons. For both systems, the soliton hopping frequency observed in a numerical simulation of a parameter sweep is well captured by the imaginary part of the eigenvalues of the Jacobian matrix, associated with the Hopf bifurcation creating the soliton hopping branch. The dynamically stable periodic orbits underlying the observed soliton hopping dynamics for both dimers and trimers are located on solution branches that originate in a Hopf bifurcation but undergo several additional saddle-node bifurcations before gaining stability. Transient oscillatory dynamics close to the Hopf bifurcation shows spatiotemporal dynamics characterized in terms of Nonlinear Dispersion Relations (NDR) ~\cite{tikan2022NonlinearDispersionRelation, leisman2019EffectiveDispersionFocusing, Lee2009Renormalized}, which can be directly related to the eigenvalues and eigenvectors associated with the Hopf bifurcation. Despite the additional saddle-node bifurcations, these characteristics associated with the Hopf bifurcation and revealed in the NDR are preserved along the solution branch and remain present within the stable periodic orbits underlying soliton hopping. 

For appropriate forcing, stationary solitons in both dimers and trimers can be mapped onto soliton solutions of a single LLE. Consequently, both solution branches exhibit foliated snaking. Within this supermode description, the periodic orbits representing dimer and trimer soliton hopping bifurcate off different segments of the soliton branch, with stability properties that are well understood in the context of snaking within a single LLE.  

The paper is organized as follows. In Sec.~\ref{sec:model}, we discuss the equations describing coupled driven-dissipative resonators, the supermode basis representation and the concept of a dispersion relation in the nonlinear regime. Further, we summarize the parametric continuation and bifurcation analysis method, the nonlinear dynamical systems approach is based on. In Sec.~\ref{sec:dimer} we present the results for the dimer system, including the bifurcation structure and descriptions of the dynamics in terms of Nonlinear Dispersion Relations (NDR). Trimer bifurcation structure is studied in Sec.~\ref{sec:trimer}. Sec.~\ref{sec:snaking} is devoted to the snaking bifurcation structure of both dimer and trimer solitons with an alternative method of forcing. The paper concludes with the discussion of the results in Sec.~\ref{sec:conclusions}.

\begin{figure*}
    \centering
    \includegraphics[width=0.9\linewidth]{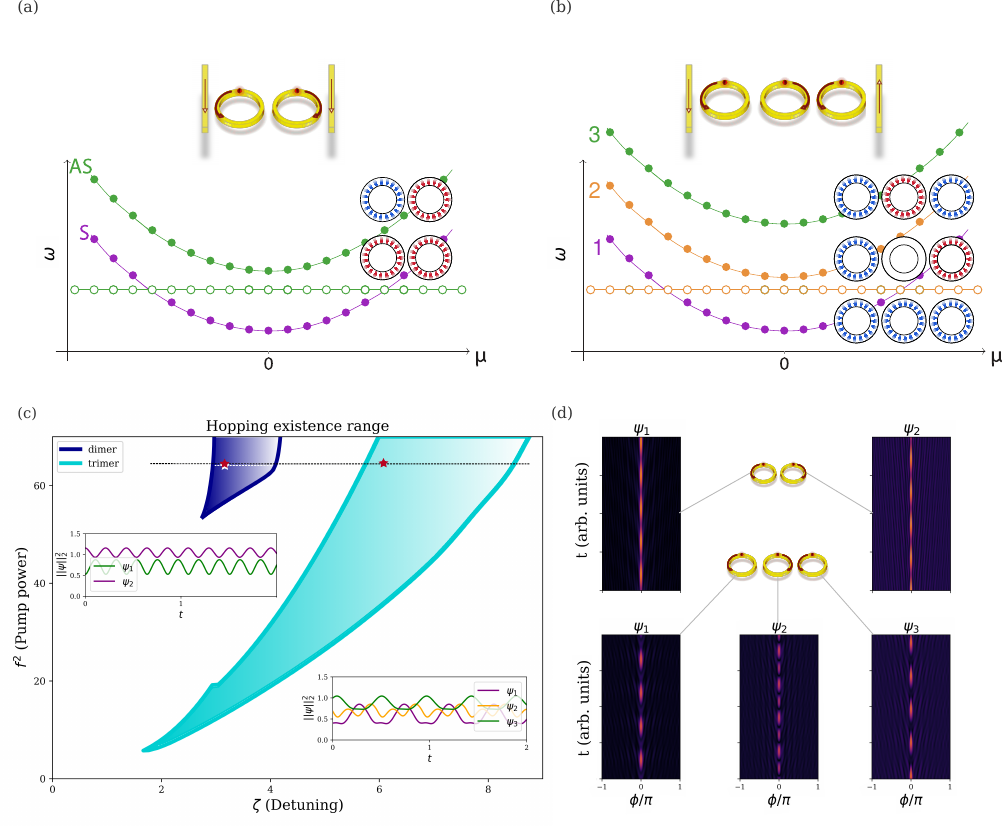}
    \caption{ \textbf{Soliton hopping in photonic molecules.} (a,b) Top: Schematic image of the photonic dimer and trimer, respectively. Arrows depict the direction of the pump. (a,b) Bottom: Corresponding schematic dispersion relation of the hybridized supermodes, where $\mu$ represents the resonator mode and $\omega(\mu)$ is the dispersion relation. The parabolic structures correspond to the dispersion relation of linear waves within different supermodes (Eq.~\ref{eq:bifhop_dispersion}), labeled by the corresponding supermode composition schematic in the resonator basis where blue and red colors indicate opposite contributions of the respective resonators in the supermode involved. A line plotted with empty dots represents a supermode soliton generated in the AS supermode of the dimer in (a) and the second supermode of the trimer in (b). (c) Boundaries of the soliton hopping existence region in the pump power ($f^2$) - detuning ($\zeta$) space as computed using the saddle-node bifurcations of the underlying time-periodic solution branches in the dimer (blue) and the trimer (cyan) for the coupling parameter $J=10$. Insets show the temporal dynamics of the intracavity field intensity of the resonators during soliton hopping in dimer and trimer  (d) Spatiotemporal dynamics of soliton hopping in the dimer (top) and the trimer (bottom).}
    \label{fig:bifhop_1}
\end{figure*}

\section{Coupled Lugiato-Lefever PDEs} \label{sec:model}

Light propagation in photonic molecules including dimers and trimers can be described by a set of coupled Lugiato-Lefever equations (LLEs)~\cite{komagata2021dissipative, Tusnin2023Nonlinear},
\begin{align}
    \partial_{t} \boldsymbol{\Psi} (\phi, t) = -(1+i\zeta)\boldsymbol{\Psi} + id_2\partial_{\phi}^2\boldsymbol{\Psi}+ i |\boldsymbol{\Psi}|^2\boldsymbol{\Psi} \nonumber \\  + i \mathcal{J}\boldsymbol{\Psi} + \boldsymbol{f}.
    \label{eq:bifhop_field}
\end{align}
We adopt the normalization introduced in Ref.~\cite{herr2014temporal}.
Here $\boldsymbol{\Psi}: [-\pi,\pi]\times \mathbb{R} \rightarrow \mathbb{C}^N = \left(\Psi_1, \Psi_2, \dots, \Psi_N\right)$ describes the slowly varying complex field envelopes in $N$ coupled cavities as a function of the angle along the ring circumference $\phi \in [-\pi , \pi]$ and time $t$, 
normalized by the photon lifetime. In line with the circular geometry, periodic boundary conditions in $\phi$ are considered. The reference frame rotates at a constant angular velocity corresponding to the group velocity of each cavity, and the angle $\phi$ is defined relative to this rotating frame. The nonlinear term $|\boldsymbol{\Psi}|^2\boldsymbol{\Psi}=\left(|\Psi_1|^2\Psi_1, |\Psi_2|^2\Psi_2, \dots, |\Psi_N|^2\Psi_N \right)$ represents the kerr nonlinearity. The parameter $d_2$ represents the normalized group velocity dispersion (assumed to be positive in this study to account for anomalous group velocity dispersion (GVD)); and 
$\zeta$ is the normalized frequency detuning between the pump and the cavity resonance, which we assume to be identical for all cavities. The normalized pumping term for each cavity is indicated by
$\boldsymbol{f} = \left(f_1, f_2, \dots, f_N\right)$ (The input pump power to each cavity $i$ is proportional to $f_i^2$), and
$\mathcal{J}$ is the normalized coupling matrix between cavities.
For a dimer and a linear chain of three coupled resonators, the trimer, the coupling matrices are given by 
\begin{equation}
    \mathcal{J}^{dimer} = \begin{bmatrix} 0 & J \\ J & 0 \end{bmatrix},\;\;\;\mathcal{J}^{trimer} = \begin{bmatrix} 0 & J & 0 \\ J & 0 & J \\ 0 & J & 0 \end{bmatrix}.
\end{equation}

Since the coupling term does not depend on the angle $\phi$ and because of periodic boundary conditions, Eq.~\ref{eq:bifhop_field} admits translations along $\phi$. The equation is equivariant under $\boldsymbol{\Psi}(\phi,t)\rightarrow\boldsymbol{\Psi}(\phi+\phi_0,t)$ where $\phi_0 \in \mathbb{R}$ corresponds to translations along the cavity angle.

It is often advantageous to consider the state of the coupled system not in the natural resonator basis but in the eigenbasis of the coupling matrix, the supermode basis. For the dimer case, linearly transforming the two intracavity fields in the eigenbasis of the coupling matrix $\mathcal{J}^{dimer}$, yields the eigenvectors (denoted as supermodes) and eigenvalues of the form 
\begin{align}
    \Psi_S &= \frac{1}{\sqrt{2}} \left(\Psi_1+\Psi_2\right), \;\;\; \lambda_S = J \nonumber \\
    \Psi_{AS} &= \frac{1}{\sqrt{2}} \left(\Psi_1-\Psi_2\right), \;\;\; \lambda_{AS} = -J.
    \label{eq:bifhop_supermode_structure_dimer}
\end{align}
In this basis, Eq.~\ref{eq:bifhop_field} can be expressed as
\begin{align}
    \partial_t \begin{bmatrix}
    \Psi_{S} \\
    \Psi_{AS}
    \end{bmatrix} =& \begin{bmatrix}
    \substack{-(1+i(\zeta-J)) \\ +id_2 \partial_\phi^2} & 0 \\
    0 & \substack{-(1+i(\zeta+J)) \\ +id_2 \partial_\phi^2}
    \end{bmatrix} \begin{bmatrix}
    \Psi_{S} \\
    \Psi_{AS}
    \end{bmatrix} \nonumber \\ 
    & + \boldsymbol{f'} + \mathcal{NL}\left( \Psi_{S}, \Psi_{AS}\right),
    \label{eq:bifhop_dimer}
\end{align}
where $\boldsymbol{f'}$ represents the normalized forcing after the transformation and $\mathcal{NL}$ represents the nonlinear part. Here, it is evident that the evolution equations for supermodes  decouple in the linear regime, and only the nonlinear part introduces cross terms introducing interactions between the two supermodes. In the supermode basis, the nonlinear terms thus control both the nonlinearity within the evolution of one supermode and the coupling between different supermodes. We refer to the supermodes as symmetric $\Psi_S$ and anti-symmetric $\Psi_{AS}$, owing to the symmetry (anti-symmetry) displayed by the cavity fields in the resonator basis. In analogy with the hybridization observed in molecular quantum mechanics, the corresponding degenerate cavity resonances undergo splitting to generate two new supermode resonances. Their frequency offsets from the cavity resonance correspond to the eigenvalues of $\mathcal{J}^{dimer}$, such that the symmetric (anti-symmetric) supermode displays resonance at $\zeta=+J(-J)$.
For the trimer case, we similarly obtain three supermodes of the form
\begin{align*}
\Psi_{S1} &= \frac{1}{2} \left(\Psi_1+\sqrt{2}\Psi_2+\Psi_3\right), \;\;\; \lambda_{S1} = \sqrt{2}J \nonumber \\
\Psi_{S2} &= \frac{1}{\sqrt{2}} \left(\Psi_1-\Psi_3\right), \;\;\;\;\;\;\;\;\;\;\;\;\;\;\; \lambda_{S2} = 0 \nonumber \\
\Psi_{S3} &= \frac{1}{2} \left(\Psi_1-\sqrt{2}\Psi_2+\Psi_3\right), \;\;\; \lambda_{S3} = -\sqrt{2}J .\nonumber \\
\end{align*}
The parabolic dispersion relation for these supermodes (displayed in Fig.~\ref{fig:bifhop_1}(a,b)(bottom)) can be better understood if we consider the following Fourier-ansatz for a solution of Eq.~\ref{eq:bifhop_dimer}
\begin{align}
    \Psi_{AS,S}(\phi, t) = \sum_\mu \Psi_{AS,S}^{\mu}(t) e^{i \mu \phi},
\end{align}
where $\mu$ represents the mode number relative to the pumped cavity mode (which is labeled as $0,\pm1,\pm2,\cdots$) and $\Psi_{AS,S}^{\mu}$ denote the complex slowly-varying modal field envelopes. By inserting this form in Eq.~\ref{eq:bifhop_dimer}, we obtain the dynamics for the symmetric supermode as
\begin{align}
    \frac{d\Psi_S^\mu(t)}{dt} = -\left(1+i\left(\zeta-J + d_2\mu^2\right)\right)\Psi_S^\mu+\delta(\mu)f'_S +\mathcal{NL},
\end{align}
where $\delta$ is the Dirac delta function, $f'_S$ is the normalized forcing for the symmetric supermode and $\mathcal{NL}$ contains the nonlinear part, which contains interactions between the modes. 

Within the linearized dynamics, the modal interactions disappear and the modal field envelope admits solutions of the form $\Psi_S^\mu(t) \propto e^{(-1+i\omega_S(\mu))t}$ (with an additional constant for the $\mu=0$ mode) leading to the following dispersion relation for the linear waves within the symmetric supermode
\begin{align}
    \omega_S(\mu) = \zeta - J + d_2\mu^2.
    \label{eq:bifhop_dispersion}
\end{align}
Similarly, for the AS supermode we obtain
\begin{align}
    \omega_{AS}(\mu) = \zeta + J + d_2\mu^2.
    \label{eq:bifhop_dispersion_2}
\end{align}
Fig.~\ref{fig:bifhop_1}(a,b)(bottom) shows the schematic of this dispersion relation of the dimer and trimer configurations along with the corresponding supermode composition. 

The presence of the nonlinear terms ($\mathcal{NL}$) in the mode equations leads to wave interactions within one supermode (intra-band) and between different supermodes (inter-band). For the cubic Kerr nonlinearity, the nonlinear term induces interactions between four modes, also called Four Wave Mixing (FWM), if those modes satisfy phase matching conditions~\cite{Chembo2010Modal} 
    \begin{align}
    \mu_1 + \mu_2 = \mu_3 + \mu_4, \nonumber \\
    \omega_1+\omega_2=\omega_3+\omega_4,
    \label{eq:bifhop_phase_matching}
\end{align}
which represent momentum and energy conservation of the interaction process.

For weak nonlinearities (which corresponds to small intracavity field intensities), the dynamics can be described by weakly coupled linear waves that follow the linear dispersion relation of the form Eq.~\ref{eq:bifhop_dispersion} on the short timescales, while on longer timescales, nonlinear interactions between the linear modes modulate their amplitudes. Interestingly, within the analogous wave turbulence approach, it has been observed that the picture of linear waves interacting via nonlinear interactions not only applies to perturbatively weak nonlinearity but also provides a good description for the dynamics with stronger nonlinearities, if the definition of the dispersion relation is modified and replaced by the Nonlinear Dispersion Relation (NDR) \cite{tikan2022NonlinearDispersionRelation}. 
The NDR is related to characteristic structures observed in the two-dimensional Fourier transform of the spatiotemporal dynamics along both space $\phi$ and time $t$, defined as 
\begin{align}
\tilde{\Psi}(\mu, \omega)=\lim_{T\to\infty}\frac{1}{T}\int_0^T\int_{-\pi}^{\pi}\Psi(\phi,t)e^{-i\mu\phi-i\omega t}d\phi dt.
\label{eq:NDR}
\end{align}
The power spectrum (PSD) $|\tilde{\Psi}(\mu, \omega)|^2$ can be used to define the NDR as the peak temporal frequencies $\omega$ in the PSD for each of the spatial modes $\mu$. The NDR can thus be interpreted as the nonlinear analog of the appropriately renormalized linear dispersion relation. For strong nonlinearities, the dynamics in a broad class of Hamiltonian wave forming systems including the Majda-McLaughlin-Tabak model~\cite{Lee2009} and the Nonlinear Schrodinger Equation~\cite{ leisman2019EffectiveDispersionFocusing} is well described by interacting weakly coupled effective linear waves that obey the renormalized, nonlinear dispersion relation. More recently, the concept of a NDR has also gained prominence in explaining complex nonlinear phenomena arising in driven dissipative counterparts of Hamiltonian systems, where the dynamics can again be described by interactions of effective wave structures~\cite{tikan2021emergent,Anderson2023}.
    
The LLE describing the optical field in a driven resonator can be interpreted as the driven-dissipative version of the Nonlinear Schrodinger Equation (NLSE), a Hamiltonian system that belongs to the class of integrable PDEs. In an infinite domain, the NLSE can be integrated using the inverse scattering transform (IST) to obtain the exact solutions in the form of \textit{radiation} (wave structures) - corresponding to the continuous spectrum of the scattering problem and \textit{solitons} (coherent structures) - corresponding to the discrete spectrum. Solitons are often understood as nonlinear solutions of the NLSE that emerge as a balance between the effects of nonlinearity and dispersion in the system. In the context of LLE, they re-emerge as ``dissipative solitons" with an additional balance between the forcing and dissipation terms. Within a description of interacting waves, dissipative solitons are viewed as the result of many cascaded FWM interactions among a broad spectrum of the cavity modes. Dictated by the double balance of nonlinearity vs. dissipation and forcing vs. dissipation, these cavity mode interactions give rise to a steady broadband \textit{$sech^2$} spectrum corresponding to the solitonic structures. Due to the broad specturm of interacting modes, the NDR of a dissipative soliton corresponds to a straight line with a slope proportional to the translation velocity of the soliton \cite{leisman2019EffectiveDispersionFocusing}.

\subsection{Nonlinear dynamical systems approach} \label{subsec:DSA}

Instead of considering spatially localized nonlinear structures within the LLE as the result of many interacting modes, one may directly consider them as solutions of the nonlinear partial differential equation, in analogy to soliton solutions in the NLSE. However, unlike in the integrable NLSE, where analytical tools such as the IST allow for the exact construction of those solutions, in the dissipative LLE exact invariant solutions underlying dissipative solitons must be computed numerically. 
This fully non-linear approach allows us to study the dynamics within coupled LLEs using dynamical systems concepts that have been successfully applied to and further developed within the context of fluid mechanics~\cite{Hopf1948, Nagataf1990, Kerswell2005, Gibson2008, Kawahara2011, Suri2017, Graham2021, Reetz2019a, Crowley2022}. Within the dynamical systems picture, the complex spatiotemporal dynamics are viewed as a trajectory in the respective system's high dimensional state space with a time evolution governed by the evolution equation. Of specific importance are so-called \emph{invariant solutions} including equilibria, representing steady evolution and periodic orbits, capturing time-periodic dynamics. Those solutions define subsets in state space, that are invariant under the dynamics and topologically correspond to points, in the case of equilibria or closed loops in the case of a periodic orbit.  Even when the system under consideration exhibits chaos, such as the turbulence observed in viscous shear flows at sufficiently high Reynolds numbers, the dynamics collapse onto a chaotic attractor, which is supported by non-chaotic yet dynamically unstable invariant solutions. Consequently, fully nonlinear invariant solutions guide the dynamics both when they are dynamically stable and thus directly observable, but also when they are unstable and support a more complex and possibly chaotic attractor. 

Steady states or equilibria $\boldsymbol{x}^*$ are the solutions of the equation $F(\boldsymbol{x}^*)=0$ where $ F(\cdot)$ is the right-hand side of the system PDEs (For this study, the coupled LLEs (Eq.~\ref{eq:bifhop_field}), $\partial_{t} \boldsymbol{\Psi} (\phi, t) = F(\boldsymbol{\Psi})$). More generally, invariant solutions both of equilibrium and periodic orbit type, satisfy the condition
\begin{align}
    \sigma \mathcal{F}^T (\boldsymbol{x}^*) - \boldsymbol{x}^* = 0,
    \label{eq:invariant}
\end{align}
 where $ \mathcal{F}^T (\cdot)$ is the time evolution operator that evolves the system state over period $T$ according to the governing equations and $\boldsymbol{x}^*$ represents either an equilibrium or a point along a periodic orbit. Note that $T$ indicates the period of a periodic orbit and is arbitrary for an equilibrium. The operator $\sigma$ represents a symmetry operator of the state, such as continuous translational symmetries. It can be used to compute invariant solutions that travel along a symmetry direction, such as relative periodic orbits of the coupled LLEs given by $\boldsymbol{\Psi} (\phi, t+T)=\boldsymbol{\Psi} (\phi+\phi_0, t)=\sigma(\phi_0)\boldsymbol{\Psi} (\phi, t)$. 
 
 Numerical computations of the invariant solutions requires an appropriate finite-dimensional representation of the state space vectors. In this study, we represent the spatiotemporal fields of the coupled cavities in a Fourier basis that naturally satisfies the periodic boundary conditions in $\phi$. The time evolution ($\mathcal{F}^T$) is computed using a $4^{th}$ order Runge-Kutta integration scheme along with implicit-explicit timestepping (We implement this step using the opensource Dedalus software~\cite{Burns2020}). The computations of invariant solutions (Eq. \ref{eq:invariant}) are performed using the opensource Channelflow library~\cite{channelflow}. This software comes with efficient Newton-Raphson methods based on matrix free Krylov subspace algorithms, such as the generalized minimal residual method (GMRES) together with hookstep trust region optimization~\cite{Saad1986, Viswanath2007}. These algorithms are capable of computing Newton steps without explicit calculations of the posssibly large Jacobian matrices. 
 
 To study the parameter dependence of these invariant solutions, we construct their bifurcation diagrams using numerical continuation tools. They are based on the predictor-corrector scheme, where cubic interpolation is used as a predictor and Newton iterations are used as corrector. Once the exact solutions are computed, their dynamic stability properties can be studied through their eigenvalue spectrum. We formulate this eigenvalue problem by considering a linear system of equations for small perturbations of the state vector, corresponding to the linearization of Eq.~\ref{eq:invariant} around the solution $\boldsymbol{x}^*$. For equilibria, the eigenvalue problem is solved using the dense eigenvalue solver routine in Dedalus. For periodic orbits, the stability properties are determined using their Floquet exponents ($\lambda$), which can be expressed through the eigenvalues $\Lambda$ of the return map of the system PDEs (Eq. \ref{eq:invariant}) using the relation $\Lambda=e^{\lambda T}$. The eigenvalues $\Lambda$ are computed using Arnoldi iterations implemented in Channelflow. 

\section{Dimer -- two coupled resonators} \label{sec:dimer}

\begin{figure*}
    \centering
    \includegraphics[width=\linewidth]{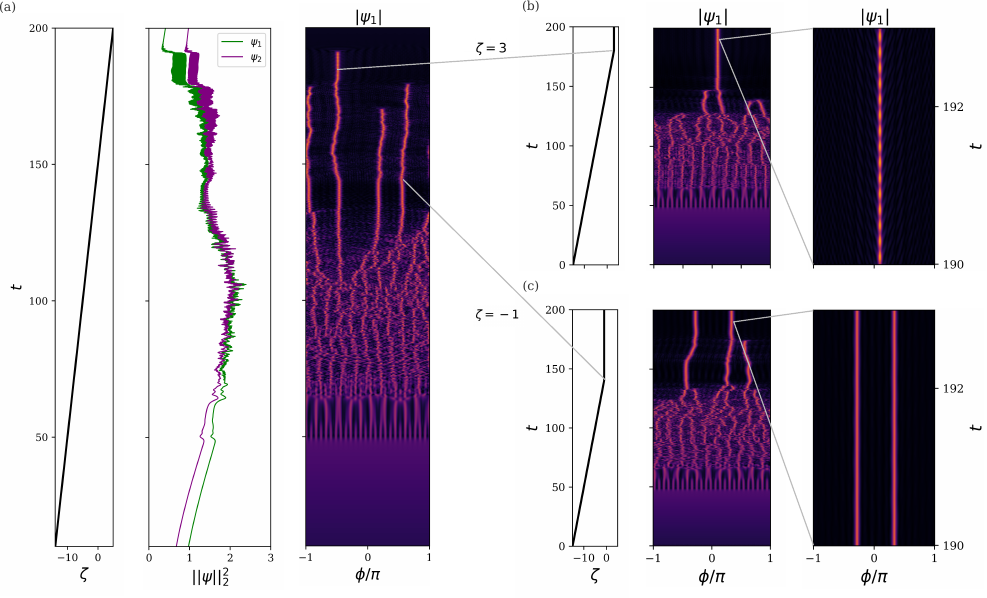}
   \caption{\textbf{Laser scanning phenomenology for the dimer}. (a) Frequency detuning scan from $\zeta=-15$ to $\zeta=5$ at a rate of $0.1$ per unit time, displaying respectively - Turing rolls, spatiotemporal chaos, supermode solitons, soliton hopping and finally continuous wave (CW) background. (b) Laser scan at the same rate is arrested at a value of $\zeta=3$, and the detuning is kept constant for the rest of the simulation. The dynamics converges to soliton hopping. (c) The ramp in the detuning is terminated at and then kept at $\zeta=-1$. The dynamics converges to the supermode soliton. These asymptotic states are then used as initial guesses for the Newton method to compute equilibrium and time-periodic solutions underlying the supermode soliton ($SS^{(2)}$) and soliton hopping ($SH^{(2)}$) respectively.}
   \label{fig:bifhop_5}
\end{figure*}

\begin{figure*}
    \centering
    \includegraphics[width=\linewidth]{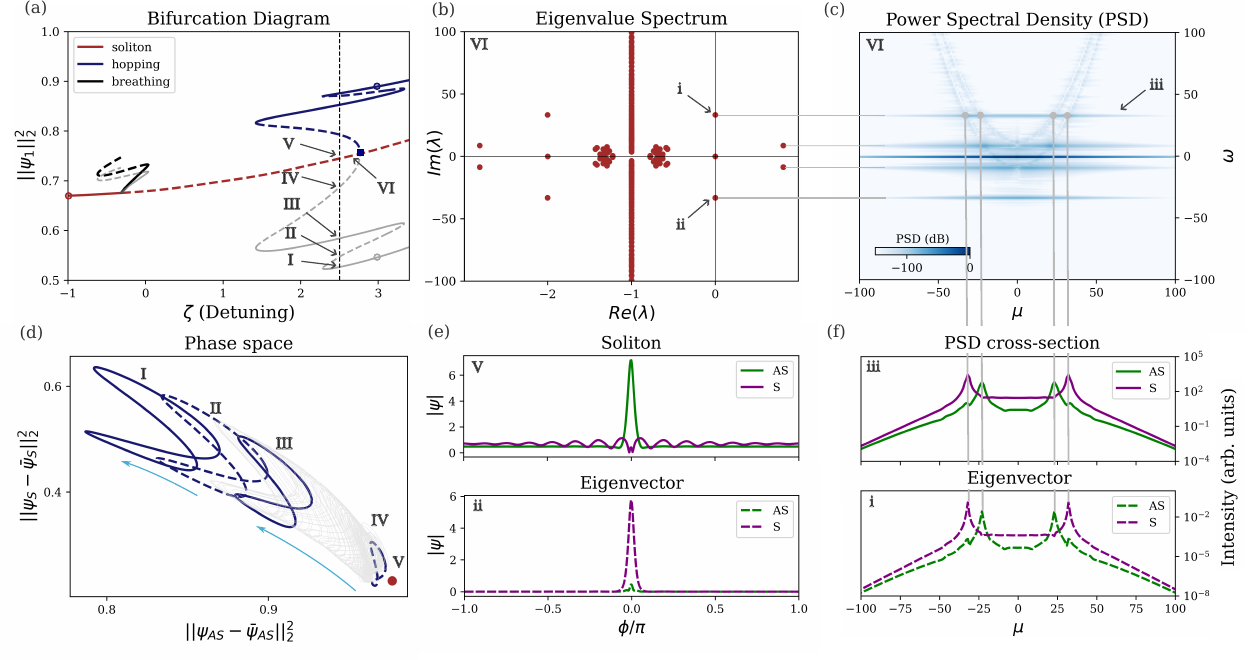}
   \caption{\textbf{Origin of dimer soliton hopping}. (a) Bifurcation diagram for the dimer system (for fixed $J=10, f=8$). Bold (dotted) lines represent stable (unstable) solutions. 
   The blue (grey) lines show maxima (minima) of the intracavity field intensity in the first resonator during one hopping period (The Hopf bifurcation corresponding to hopping marked with $\square$). The bifurcation branches are constructed using asymptotic states of laser scans (Fig.~\ref{fig:bifhop_5}) terminated at $\zeta=-1$ for soliton and $\zeta=3$ for soliton hopping as initial guesses for the Newton method (Marked with $\circ$)
   (b) Eigenvalue spectrum of $SS^{(2)}$ close to bifurcation at $(VI)$, featuring a complex conjugate pair of eigenvalues on the imaginary axis that is responsible for the Hopf bifurcation.
   (c) PSD of the spatiotemporal dynamics representing the transient nonlinear dispersion relation for the pumped resonator. 
   The system is initialized at $SS^{(2)}$ close to bifurcation at $(VI)$ along with small random perturbations of magnitude $||\delta \Psi||_2 = 10^{-3}$ and evolved for $T=10$.
   (d) Phase space projection of the spatiotemporal fields of various solutions present at $\zeta=2.5$. 
   The grey line displays a trajectory of the system initialized at an unstable $SH^{(2)}$ $(IV)$ with small random perturbations. The system eventually converges to a stable $SH^{(2)}$ $(III)$.
   (e) Top: Spatial profile of $SS^{(2)}$ $(V)$, displaying a solitonic structure in the AS supermode. (e) Bottom: Spatial profile of the eigenvector $(ii)$, displaying a solitonic structure in the S supermode.
   (f) Top: Cross section of the PSD in (c) at $\omega=33.27$ in the supermode basis. (f) Bottom: Power spectrum of the eigenvector $(i)$. The eigenvector displays similar profile to the PSD cross section at the frequency that corresponds to its eigenvalue.
   }
   \label{fig:bifhop_2}
\end{figure*}

We begin our analysis by considering the dimer configuration under the influence of 
pump forcing coupled to the first resonator. To maintain consistency with microresonator fabrication parameters, 
we set the dispersion parameter to $d_2 = 0.04$. 
The phenomenon of soliton hopping has been observed in the strongly coupled regime~\cite{tikan2021emergent}, requiring a high minimum threshold value of the coupling parameter~\cite{komagata2021dissipative}. 
To investigate this phenomenon, we therefore set the coupling parameter to the value of $J = 10$, 
sufficient to observe soliton hopping in both dimer and trimer configurations. We fix the pumping term for both dimer and trimer to $f^2=64$. 

With these parameter settings, we study the spatiotemporal dynamics of the dimer by 
scanning through the normalized detuning parameter $\zeta$ over the AS supermode resonance (which occurs at $\zeta_{AS}=-J=-10$ (Eq.~\ref{eq:bifhop_dispersion_2}))~\cite{herr2014temporal}. The laser scan represents a typical experimental setting for studying microresonator devices, where the pump laser frequency is varied at a constant rate close to the cavity resonances. Fig.~\ref{fig:bifhop_5}(a) shows such a simulation where the system is initialized with random noise and $\zeta$ is varied from $-15$ to $5$ at a constant rate of $0.1$ per unit normalized time.
This exploration unveils a sequence of 
intriguing phenomena, well-documented in the context of coupled resonator systems \cite{tikan2021emergent, komagata2021dissipative}, 
which include the formation of Turing rolls, spatiotemporal chaos, the emergence of supermode solitons, and ultimately, the onset of soliton hopping.

 Fig.~\ref{fig:bifhop_5}(b) displays a simulation in which $\zeta$ is ramped up at the same rate until $\zeta=-1$, where it is kept constant for the rest of the simulation. In this case the dynamics  converges to a solitonic structure. Similarly, terminating the ramp at $\zeta=3$ leads to the dynamics converging to time-periodic soliton hopping (Fig.~\ref{fig:bifhop_5}(c)). Using those asymptotic system states as initial guesses for Newton's method we obtain an equilibrium solution $SS^{(2)}$ corresponding to the solitonic structure at $\zeta=-1$, and a periodic orbit solution $SH^{(2)}$ corresponding to soliton hopping at $\zeta=3$. These are two converged invariant solutions of the nonlinear governing equations for dimer (Eq.~\ref{eq:bifhop_dimer}). To understand the relationship between these two invariant solutions, we follow their solution branches in $\zeta$ and construct bifurcation diagrams using numerical continuation methods (Fig.~\ref{fig:bifhop_2}(a)). The bifurcation analysis shows that $SH^{(2)}$ emerges through a subcritical Hopf bifurcation from $SS^{(2)}$ at a critical value of $\zeta=2.767$.

Fig.~\ref{fig:bifhop_2}(b) shows the eigenvalue spectrum of $SS^{(2)}$ near the bifurcation point $(VI)$. In this spectrum, the neutral eigenvalue $\lambda=0$ corresponds to the translational symmetry of the governing equations along $\phi$. All eigenvalues are either real or a part of a complex conjugate pair. They are symmetrically located about the $Re(\lambda)=-1$ line, which is related to the Hamiltonian nature of the augmented linear problem \cite{Qi2019}. We observe a continuous spectrum of eigenvalues located at $Re(\lambda)=-1$ which correspond to dispersive waves (DW) on the continuous wave (CW) background of the supermode soliton (i.e. the homogeneous background). Additionally, we identify two complex conjugate eigenvalues at $Re(\lambda)=0.811$ that are related to soliton breathing. They give rise to an intermediate periodic orbit solution branch at $ \zeta'=-0.314 $ that exists within a short window ($\zeta \in \left[-0.314, 0.036\right]$) and becomes unstable after a saddle-node bifurcation. Notably, during the laser scanning experiments, this breathing structure is not observed, presumably because the laser detuning parameter $\zeta$ is varied at a finite rate and thus not quasi-statically. When the rate of change of the detuning parameter is large compared to the growth rate associated with the oscillatory instability of $SS^{(2)}$, the system `jumps' over the detuning window, in which the $SS^{(2)}$ is unstable and instead reaches soliton hopping. Of particular interest are two complex conjugate eigenvalues that cross the imaginary axis (Fig.~\ref{fig:bifhop_2}(b)(i,ii)). These eigenvalues are responsible for the bifurcation of the $SH^{(2)}$ branch from the $SS^{(2)}$ branch.

After its origin at the subcritical Hopf bifurcation, the $SH^{(2)}$ branch undergoes multiple saddle-node or fold bifurcations in $\zeta$, leading to the existence of multiple periodic orbits capturing soliton hopping in the same range of the detuning parameter. The Floquet multipliers of these multiple solution branches reveal that they are alternatingly stable and unstable (Fig.~\ref{fig:bifhop_2}(a)). At the Hopf bifurcation, the $SH^{(2)}$ branch is initially unstable. To highlight the stability features of these branches, Fig.~\ref{fig:bifhop_2}(d) shows a phase space plot of the invariant solutions at a fixed parameter value of $\zeta=2.5$. All the invariant solutions found at this $\zeta$ are projected onto a 2-dimensional space based on the $L_2$ norm of the deviations of intracavity fields from the CW background in the supermode basis. This phase space plot depicts four solutions on the $SH^{(2)}$ branch: two unstable (dashed, $(II)$ and $(IV)$) and two stable (solid, $(I)$ and $(III)$) — that become increasingly distinct from $SS^{(2)}$ $(V)$ as we move away from the $SS^{(2)}$ branch on the $\zeta=2.5$ line of the bifurcation diagram. When we initiate the system dynamics close to an unstable $SH^{(2)}$ $(IV)$ with small random perturbations, it eventually converges towards the subsequent stable $SH^{(2)}$ ($(III)$, the trajectory depicted in grey).

To investigate the relevance of these invariant solutions and their bifurcation structure for the observed dynamics, we analyze the spatiotemporal evolution of the system close to the bifurcation point $\zeta=2.767$. Here we choose to depict the dynamics in Fourier space, using the power spectral density (PSD) of the fields in space $\phi$ and time $t$. The high intensity patterns in this PSD can be viewed as the NDR of the system (as defined by Eq.~\ref{eq:NDR}). In Fig.~\ref{fig:bifhop_2}(c), we present the PSD for the spatiotemporal field of the forced resonator when the dimer system is initialized at $SS^{(2)}$ $(VI)$ with random perturbations of magnitude $||\delta \Psi||_2 = 10^{-3}$, and followed for an integration time of $T=10$.  Note that the data is collected for the transient dynamics, which remains in the vicinity of $SS^{(2)}$ so that the linearization around this state remains relevant. 

\begin{figure*}
    \centering
    \includegraphics[width=0.8\linewidth]{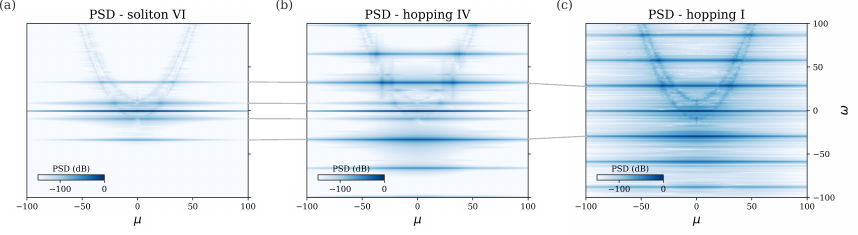}
   \caption{\textbf{PSD evolution along the solution branch}. Left: PSD of the spatiotemporal dynamics of $SS^{(2)}$ close to the bifurcation point $(VI)$ (same as Fig.~\ref{fig:bifhop_2}(c)). Center and right: PSD of the hopping dynamics. The system in initialized at the $SH^{(2)}$ $(IV)$ and $(I)$ at $\zeta=2.5$ with small random perturbations of magnitude $||\delta \Psi||_2 = 10^{-3}$ and evolved for $T=10$. Some features of the PSD close to the Hopf bifurcation $(VI)$ are retained in the PSD for stable $SH^{(2)}$ $(I)$ despite the multiple saddle-node bifurcations along the solution branch.
   }
   \label{fig:bifhop_6}
\end{figure*}

In this PSD plot, the horizontal line at $\omega=0$ represents the spatially localized stationary supermode soliton. Transforming the PSD into the supermode basis, which implies a linear transformation of the PSD data in both resonators using Eq.~\ref{eq:bifhop_supermode_structure_dimer}, reveals that this soliton predominantly exists within the AS supermode~\cite{komagata2021dissipative}, and is consequently called \emph{supermode soliton}. The random initial perturbations excite additional modes beyond the unstable $SS^{(2)}$ state in the transient dynamics. These appear as additional features in the PSD plot.  Dispersive waves (DW) obeying the linear dispersion relation ~\ref{eq:bifhop_dispersion} for the CW state are represented by two parabolic structures shifted by twice the inter-resonator coupling $J$. Time-periodic structures are dominated by temporal Fourier mode corresponding to the temporal period and thus manifest themselves in the PSD as a pair of horizontal lines with spacing equal to the twice the repetition frequency $\omega_p$. They appear as horizontal lines, because their spatial localization implies a broad Fourier spectrum in $\mu$. 
The spacing of these horizontal lines in the PSD plot corresponds to the complex conjugate eigenvalues with $Im(\lambda)=\pm \omega_p= \pm 33.27$ associated with the Hopf bifurcation, as seen in the comparison between Fig.~\ref{fig:bifhop_2}(b) and (c) (see gray connecting lines). Further, the power spectrum of the associated eigenvectors accurately captures the cross sections of PSD at the frequencies corresponding to their eigenvalues (Fig.~\ref{fig:bifhop_2}(f)). 
Likewise, we also observe the breathing instability of $SS^{(2)}$ in the form of two horizontal lines at the frequencies that correspond to the leading complex conjugate eigenvalues with $Im(\lambda)\approx\pm 8.67$ associated with breathing. Consequently, the PSD of the transient dynamics in the vicinity of $SS^2$ well captures the linearized dynamics as described by the eigenvalue spectrum and the corresponding eigenvectors. 

The PSD plot with its interpretable geometric features suggests to describe the spatio-temporal dynamics in terms of wave structures that evolve according to their specific dispersion relation and exchange energy via nonlinear interactions. Here, the dispersion relation of the nonlinear waves is given by their NDR and wave interactions are controlled by phase matching conditions Eq.~\ref{eq:bifhop_phase_matching} encoding the four-wave-mixing processes (FWM) induced by the cubic nonlinearity. In the supermode basis, these processes can be further classified into inter-band and intra-band processes based on the interactions within one supermode and interactions between different supermodes, as described in detail by ~\cite{komagata2021dissipative}. Within this picture, the emergence of soliton hopping has been explained via the interactions between AS and S supermodes such that the AS supermode soliton (Fig.~\ref{fig:bifhop_2}(e)(top)) excites a soliton in the S supermode. Since the S supermode soliton has a frequency shift $\omega_p$ relative to the AS supermode, the periodic beating of the two supermode solitons then gives rise to soliton hopping. Accordingly, we find that one of the eigenvectors associated with the Hopf bifurcation that generates $SH^{(2)}$ is indeed a solitonic structure in the S supermode (Fig.~\ref{fig:bifhop_2}(e)(bottom)). 

Key features of the PSD for $SH^{(2)}$ close to the bifurcation point are preserved further along the solution branch, even beyond three saddle-node bifurcations that give rise to two stable $SH^{(2)}$ periodic orbits coexisting at the same detuning parameter value. Fig.~\ref{fig:bifhop_6} shows the PSD close to the bifurcation from the $SS^{(2)}$ indicating dispersive waves, the breathing instability and the interacting supermode solitons underlying soliton hopping, together with the PSDs for the unstable periodic obit at $(IV)$ and the stable periodc orbit underlying robust soliton hopping at $(I)$. The detuning for both periodic orbits is $\zeta=2.5$ and PSDs have been computed for simulations initiated with the perturbation magnitude $||\delta \Psi||_2 = 10^{-3}$ and followed over a time interval of $T=10$. 
The horizontal lines associated with the Hopf bifurcation are preserved, and the hopping frequency only shifts slightly (see grey lines). The PSD at $(I)$ in addition shows an entire ladder of horizontal lines indicating harmonics of the hopping frequency. In accordance with the stabilization via saddle-node bifurcations, the lines associated with the breathing instability subside as we march along the branch from IV to I. Even in the fully developed soliton hopping situation, dispersive waves remain present in the dynamics. 
The stable periodic orbit underlying the robust soliton hopping phenomenon observed in simulations of laser scans thus retains key features that can be explained in terms of the Hopf bifurcation off the $SS^{(2)}$ branch creating the $SH^{(2)}$ branch, even though the branch undergoes several additional bifurcations before the stable periodic orbit is reached. This suggests that the mode couplings creating the temporal beating between two detuned supermode solitons remain robust. However, due to the additional saddle-node bifurcations that eventually stabilize the solution, a formal weakly nonlinear description of the saturated stable periodic orbit, which would be appropriate for points on the periodic orbit branch in the direct vicinity of the initial Hopf bifurcation, does not appear straightforward.

\section{Trimer -- three coupled resonators} \label{sec:trimer}

\begin{figure*}
    \centering
    \includegraphics[width=\linewidth]{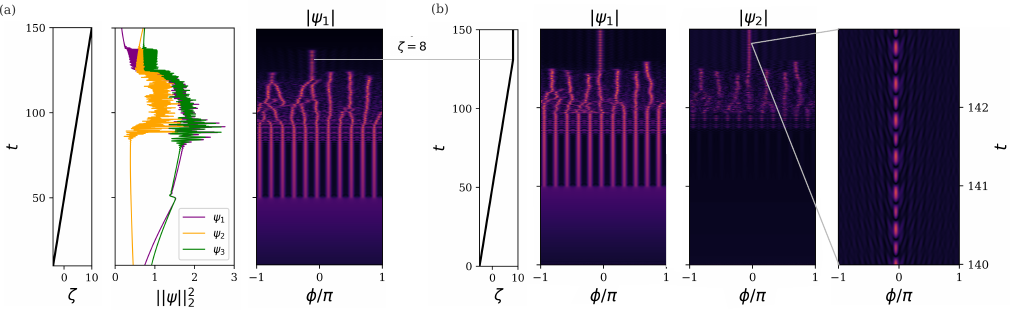}
   \caption{\textbf{Laser scanning phenomenology for the trimer}. (a) Frequency detuning scan from $\zeta=-5$ to $\zeta=10$ at a rate of $0.1$ per unit time, displaying respectively - Turing rolls, spatiotemporal chaos, soliton hopping and finally continuous wave (CW) background. (b) Laser scan is swept until $\zeta=8$ at the same rate, where it is kept constant for the rest of the simulation. The system dynamics converges to the soliton hopping structure. This asymptotic state is used as an initial guess for the Newton method to compute the underlying time-periodic solution $SH^{(3)}$}
   \label{fig:bifhop_7}
\end{figure*}

\begin{figure*}
    \centering
    \includegraphics[width=0.7\linewidth]{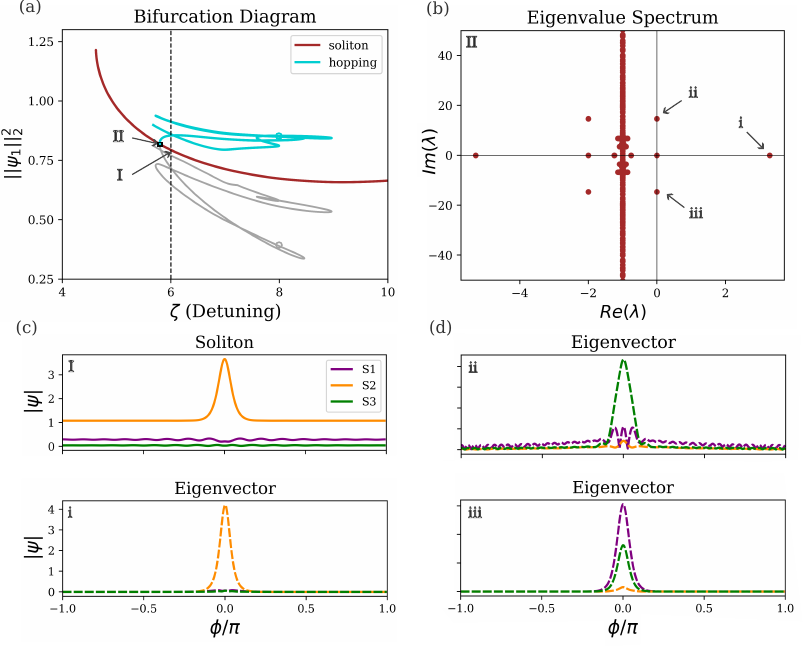}
   \caption{\textbf{Origin of trimer soliton hopping}
   (a) Bifurcation diagram for the trimer system (for fixed $J=10, f=8$) (Stability features are not shown). The soliton hopping solution branch was computed using the asymptotic state obtained during the laser scan (Fig.~\ref{fig:bifhop_6}) terminated at $\zeta=8$ as an initial guess for the Newton method (marked with $\circ$). Since the supermode soliton was not observed during the laser scans the soliton solution branch was constructed using the soliton hopping solution close to the Hopf bifurcation as an initial guess.
   (b) Eigenvalue spectrum of $SS^{(3)}$ close to the bifurcation $(II)$, displaying a complex conjugate pair of eigenvalues on the imaginary axis that is responsible for the Hopf bifurcation (c) Top: Spatial profile of $SS^{(3)}$ $(I)$. (c) Bottom: Spatial profile of the eigenvector $(i)$. Both $(I)$ and $(i)$ display a solitonic structure in the S2 supermode.
   (d) Spatial profiles of eigenvectors top: $(ii)$ and bottom: $(iii)$, associated with the Hopf bifurcation. Both eigenvectors have solitonic profiles predominantly in the S1 and S3 supermodes.
   } 
   \label{fig:bifhop_3}
\end{figure*}

We consider a trimer consisting of a linear chain of three coupled resonators subject to forcing of the first resonator. The individual resonator parameters are chosen as in the dimer case, with dispersion parameter $d_2=0.04$. Likewise, the coupling between two neighboring resonators, $J=10$, and the forcing amplitude, $f^2=64$, are unchanged compared to the dimer. Fig.~\ref{fig:bifhop_7}(a) shows a laser detuning scan over the S2 supermode resonance as $\zeta$ is varied from -5 to 10 at a constant rate of 0.1 per unit time. 
During the scan, the trimer system shows first the formation of Turing rolls, then spatiotemporal chaos, soliton hopping, and eventually decay to the continuous wave (CW) background state. Unlike for the dimer case, soliton hopping in the trimer system is not preceded by a stationary supermode soliton state. Here soliton hopping takes the form of a solitonic structure jumping from one resonator to its neighbor and being reflected at the open ends of the chain (Fig.~\ref{fig:bifhop_1}(d)(bottom)). For the trimer, the frequency of soliton oscillations observed in the middle resonator is thus twice the frequency of those observed in the two outer resonators. 

Fig.~\ref{fig:bifhop_7}(b) shows a laser scan where $\zeta$ is ramped up at the same rate but arrested at $\zeta=8$, and kept constant for the rest of the simulation. Now the soliton hopping dynamics persists and can be used as an initial guess for a Newton search. The search converges to a periodic orbit solution $SH^{(3)}$ underlying soliton hopping at $\zeta=8$. Starting from this periodic orbit solution, we numerically continue the solution branch as a function of the the detuning parameter. The bifurcation diagram, Fig.~\ref{fig:bifhop_3}(a),  reveals that the $SH^{(3)}$ soliton hopping branch bifurcates from an equilibrium supermode soliton $SS^{(3)}$ in a supercritical Hopf bifurcation at $\zeta=5.79$ $(II)$. However, after the initial Hopf bifurcation, the branch of periodic orbits undergoes an additional saddle-node bifurcation before reaching the $SH^{(3)}$ solution at $\zeta=8$ that is observed in the laser scan. Continuing the $SS^{(3)}$ branch from the bifurcation point $(II)$ completes the bifurcation diagram. 

To characterize the Hopf bifurcation, we compute the eigenvalue spectrum of $SS^{(3)}$ equilibrium at the bifurcation point, shown in Fig.\ref{fig:bifhop_4}(b). The neutral modes of the Hopf bifurcation given by the eigenvectors associated to the complex conjugate pair of eigenvalues crossing the imaginary axis display soliton structures in the S1 and S3 supermodes (Fig.~\ref{fig:bifhop_3}($ii$, $iii$)). In contrast, the supermode soliton branch $SS^{(3)}$ itself predominantly resides in the S2 supermode. The excitation of S1 and S3 supermodes in the Hopf bifurcation can also be interpreted within the wave interaction picture. Via the FWM process, the S2 supermode soliton projects onto solitonic structures in the S1 and S3 supermode. These S1 and S3 solitons have a frequency offset with respect to the S2 soliton leading to temporal beating and thereby periodic soliton hopping. 
While the $SH^{(3)}$ is created in a standard supercritical Hopf bifurcation, the $SS^{(3)}$ branch it bifurcates from is already unstable, as evidenced by a real positive leading eigenvalue. The associated eigenvector resides in the S2 supermode as has the shape of the supermode soliton indicating instability with respect to the amplitude of the $SS^{(3)}$ equilibrium solution. The bifurcating $SH^{(3)}$ branch inherits this instability and is only stabilized via an additional saddle-node bifurcation. This bifurcation scenario of a supercritical Hopf bifurcation off an unstable equilibrium branch is compatible with the fact, that the supermode soliton state is not observed during a laser scan and instead, soliton hopping supported by the stable periodic orbit directly follows the spatiotemporal chaos. 

\section{Snaking in supermode-forced dimers and trimers} \label{sec:snaking}

For both the dimer and the trimer configuration, we have identified periodic orbits underlying soliton hopping, which both bifurcate in Hopf bifurcations from respective stationary supermode solitons. These spatially localized equilibrium solutions of coupled systems of LLEs predominantly reside in one supermode (for the dimer in the asymmetric supermode AS, and for the trimer in the second supermode S2), but also excite weak dispersive waves in the other supermodes. To eliminate the dispersive wave components and map supermode solitons in dimers and trimers exactly onto soliton solutions of a single LLE, we modify the forcing protocol so that only one supermode is forced. 

Consider the dimer configuration in the supermode basis governed by Eq.~\ref{eq:bifhop_dimer} with a forcing $\boldsymbol{f'}=(f_S,f_{AS})=(0,f)$. In the resonator basis this forcing corresponds to $\boldsymbol{f}=\left(\frac{f}{\sqrt{2}},-\frac{f}{\sqrt{2}}\right)$. 
For such forcing, Eq.~\ref{eq:bifhop_dimer} is homogeneous for the symmetric supermode component and admits solutions of the form $(\Psi_{S},\Psi_{AS})=(0, \Psi_0)$ where $\Psi_0$ satisfies the single LLE 
\begin{align}
     \partial_{t} \Psi_0 (\phi, t) &= -(1+i\tilde{\zeta})\Psi_0 + id_2\partial_{\phi}^2\Psi_0
         \nonumber \\ 
         &+ i |\Psi_0|^2\Psi_0 + f.
         \label{eq:bifhop_mapping}
\end{align}
Here, $\tilde{\zeta}=\zeta+J$ is the adjusted detuning based on the location of the AS supermode resonance. Likewise, for the trimer configuration, an S2 supermode forcing $\boldsymbol{f'}=(0,f,0)$, corresponding to $\boldsymbol{f}=\left(\frac{f}{\sqrt{2}},0,-\frac{f}{\sqrt{2}}\right)$ in the resonator basis, allows for solutions of the form $(\Psi_{1},\Psi_{2},\Psi_{3})=(0, \Psi_0,0)$ with $\Psi_0$ satisfying the single LLE for $\tilde{\zeta}=\zeta$. 

Despite the modified forcing, both the dimer and trimer show essentially the same dynamical regimes that are observed for the single resonator forcing case. Even for experimentally easier to achieve coupling and forcing, with parameter values $J=7$ and $f^2=16$, the two systems show soliton hopping. For laser scans with increasing detuning parameter the sequence of observed regimes also does not differ between the two forcing protocols. Particularly, soliton hopping for a supermode-forced dimer is preceded by the observation of a stationary supermode soliton, while the supermode-forced trimer shows the onset of soliton hopping without prior emergence of a stationary soliton.  

As before, converging guesses extracted from the dynamics using Newton iterations, we can compute periodic orbit solutions $SH^{(2')}$ and $SH^{(3')}$ that underlie the soliton hopping in the dimer and trimer with the supermode forcing. Obviously, the solutions differ from the previously discussed $SH^{(2)}$ and $SH^{(3)}$ solutions for a single cavity forcing, but they display very similar spatial and temporal features, indicating that the soliton hopping phenomenon primarily arises from the coupling structure between the cavities and is robust against different types of forcing. Parametric continuation shows that the periodic orbit branches of both dimers and trimers bifurcate from a stationary soliton branch that is entirely confined to the forced supermode. Consequently, the equilibrium branch that both periodic orbit branches bifurcate from represents a pure supermode soliton without additional DW contributions and can be exactly mapped to solutions of the single LLE.

For the single LLE defined on a periodic domain, the bifurcation structure of several types of spatially localized equilibrium solutions has been studied in detail \cite{parra-rivas2018BifurcationStructureLocalized, Parra-Rivas2016, Barashenkov1998Bifurcation}. For low values of detuning ($\tilde{\zeta}<2$), solution branches of spatially localized patterns bifurcate from the periodic Turing roll pattern branch, at a detuning value extremely close to the crictical value, where the modulational instability causes the Turing roll branch itself to bifurcate off the unpatterned homogenous CW background.  
The spatial structure of these localized patterns is characterized by a finite patch of Turing roll pattern embedded in the homogeneous background. Solutions of different spatial extent or different number of Turing rolls coexist at the same detuning values and exhibit spatially oscillating tails connecting the patterned interior to the unpatterned background state. Those solutions are arranged in a classical snakes-and-ladders bifurcation structure resembling the snaking solutions of the 1D Swift-Hohenberg equation with 2-3 nonlinearity, where both the emergence of spatial localization and the growth of solutions under parametric continuation has been studied in detail~\cite{Burke2006}. 

\begin{figure*}
    \centering
    \includegraphics[width=\linewidth]{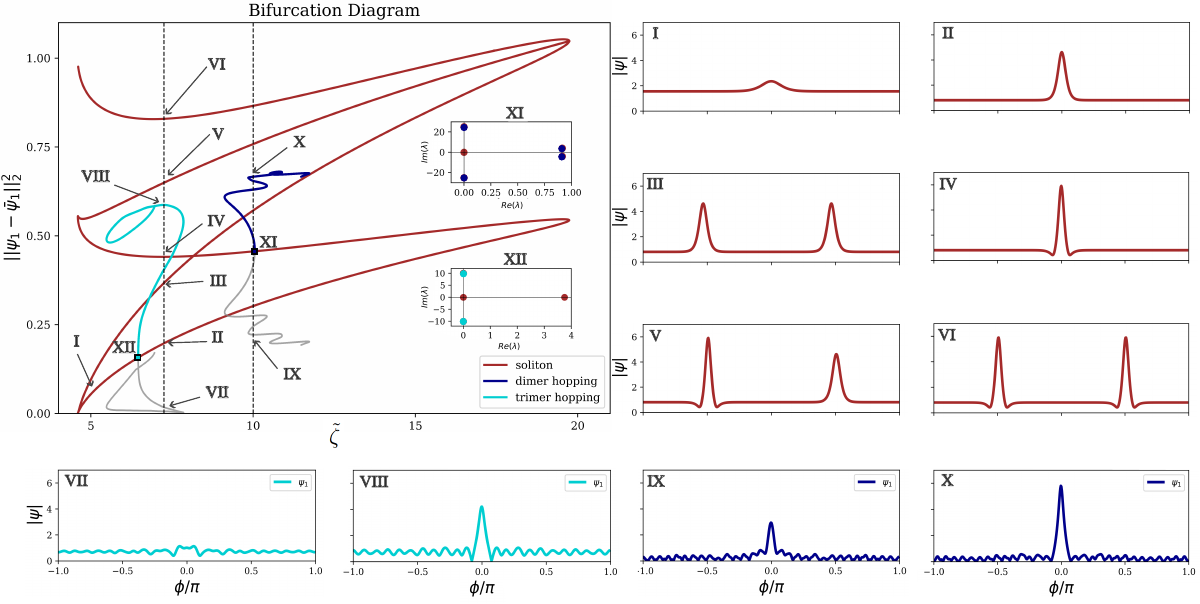}
   \caption{\textbf{Foiliated snaking of supermode solitons.}
   Top left: Bifurcation diagram for both dimer and trimer systems with a supermode forcing. x-axis: $\tilde{\zeta}$ is the adjusted detuning based on the location of the supermode resonance, 
   $\tilde{\zeta} = \zeta + J$ for AS supermode in dimer and $\tilde{\zeta}=\zeta$ for S2 supermode in trimer. 
   y-axis: $||\psi_1 - \bar{\psi}_1||^2_2$ - Intracavity field intensity in the first resonator after removing the constant CW background (Hopf bifurcations corresponding to hopping marked with $\square$). Insets $XI$ and $XII$ show the eigenvalue spectrum of dimer and trimer supermode solitons ($SS^{(2')}$ and $SS^{(3')}$) respectively at the corresponding Hopf bifurcations. The eigenvectors corresponding to \textcolor{brown}{$\bullet$} have components exclusively in the forced supermode and represent instabilities inherited from the single LLE. The eigenvector corresponding to $\lambda=0$ is associated with the continuous translational symmetry of the system. Eigenvectors corresponding to \textcolor{darkblue}{$\bullet$} and \textcolor{darkturquoise}{$\bullet$} have non-zero field components in the unforced supermodes of the dimer and the trimer respectively, and represent emergent instabilities arising from the coupling of the supermodes.
   \textbf{$I-VI$}: Spatial profiles of the intracavity field for the supermode soliton solution (AS for dimer and S2 for trimer) at different locations of the bifurcation diagram.
    \textbf{$VII-X$}: Spatial profiles of the intracavity field in the first resonator ($\psi_1$) at different moments during the hopping dynamics corresponding to the extrema of the observable - dimer (cyan) and trimer (blue).
   }
   \label{fig:bifhop_4}
\end{figure*}

The soliton solutions with monotonic, non-oscillating spatial tails considered here, are not part of the previously discussed localized solutions but directly bifurcate off the homogenous background at higher values of detuning ($\tilde{\zeta}>2$). This bifurcation occurs close to the detuning parameter value at which the homogenous background itself undergoes a saddle-node bifurcation, creating a region sometimes termed `bistable'~\cite{Parra-Rivas2014} because within the subspace of homogenous, e.g. translationally invariant solutions, there are CW solutions with 3 different amplitudes of which 2 are stable with respect to homogenous perturbations. However, in the context of the PDE describing spatial variation, the upper branch is unstable with respect to spatially varying eigenmodes so that the description as `bistable' in the context of anomalous dispersion is misleading. Localized single solitons as well as multi-peak solitons comprising weakly interacting isolated peaks exist for the same parameter values. These soliton solutions are organized within a \emph{foliated snaking} bifurcation structure. 
For the forcing parameter $f^2=16$ considered in this section, these soliton solution branches bifurcate from the homogeneous background at $\tilde{\zeta}=4.599$ and exist within the range of detuning values $\tilde{\zeta} \in [4.599,19.739]$. The foliated bifurcation structure of the corresponding supermode soliton solutions of the supermode-forced dimer ($SS^{(2')}$) and trimer ($SS^{(3')}$) is shown in Fig.~\ref{fig:bifhop_4}, where we plot the intracavity field intensity in the first resonator after removing the constant CW background.

Having mapped the stationary soliton solutions in both dimers and trimers onto soliton solution of a single LLE, we can now also represent the soliton hopping solution branches of both the supermode-forced dimer and the trimer systems in this one bifurcation diagram. Note, however, that the periodic orbit branches shown in Fig.~\ref{fig:bifhop_4}, exist only in either the dimer, or the trimer system. The $SH^{(2')}$ solution branch bifurcates off the $SS^{(2')}$ branch at $\tilde{\zeta}=10.05$ in a subcritical Hopf bifurcation, mimicking the emergence of the $SH^{(2)}$ branch in the single-resonator forced dimer case. Likewise, the $SH^{(3')}$ solution branch bifurcates from the $SS^{(3')}$ branch at $\tilde{\zeta}=6.46$ in a supercritical Hopf bifurcation, again mimicking the type of bifurcation, by which the $SH^{(3)}$ is formed in the single-resonator forced trimer case. The eigenvalue spectra of $SS^{(2')}$ and $SS^{(3')}$ solutions at these Hopf bifurcations (Fig.~\ref{fig:bifhop_4}(Top left: inset)) are similar to the single resonator forcing case, with a complex conjugate pair of leading unstable eigenvalues for $SS^{(2')}$ (similar to $SS^{(2)}$ spectrum) and a leading unstable real eigenvalue for $SS^{(3')}$ (similar to $SS^{(3)}$). 

The instabilities of the stationary supermode soliton solutions $SS^{(2')}$ and $SS^{(3')}$, can be classified into two categories based on the eigenvectors associated with the unstable eigenvalues: instabilities with eigenvectors that are entirely confined to the forced supermode and instabilities with eigenvectors components in the unforced supermodes. The former corresponds to instabilities of the soliton solutions of the single LLE. The latter results from the coupling between the forced supermode and the unforced supermodes; those instabilities only exist in the dimer and trimer systems and are absent in a single resonator. 
For both the dimer and the trimer, the Hopf bifurcation creating the periodic orbit branch underlying soliton hopping is associated with neutral eigenvectors of the equilibrium branches $SS^{(2')}$ and $SS^{(3')}$ that have non-zero overlap with the unforced supermodes. Consequently, these Hopf bifurcations do not exist in a single LLE, but only emerge in a coupled resonator system. This corroborates the interpretation of soliton hopping as an emergent phenomenon due to the resonator-resonator coupling induced by nonlinearity.  

As for the single resonator forcing, the periodic orbits underlying soliton hopping in the supermode-forced dimer and trimer bifurcate from equilibrium branches that are already unstable. However, the nature of the instability of the equilibrium branch differs and can be inferred from the eigenvectors associated with the unstable eigenvalues. For the dimer, at the Hopf bifurcation creating the $SH^{(2')}$ branch, the spectrum of the soliton equilibrium branch $SS^{(2')}$ shows an unstable leading pair of complex conjugated eigenvalues with associated eigenvectors that are exclusively contained in the unforced supermode. Consequently, this oscillatory instability only exists in the dimer system, while the equivalent soliton branch in a single resonator described by a single LLE is stable. For the trimer, however, the $SH^{(3')}$ bifurcates off $SS^{(3')}$, where the equilibrium branch has a single unstable real-valued eigenvalue with an eigenvector entirely contained in the forced supermode. This instability is consequently also present in the single LLE and thus `inherited' from the single resonator soliton. The fact that the trimer supermode soliton at the onset of soliton hopping is unstable to an instability within the same forced supermode while the dimer supermode soliton is only unstable with respect to an instability in the other supermode may help rationalize, why observing the supermode soliton in a laser scan with finite rate of change of the detuning may be easier in a dimer than in a trimer. Note however, that the characteristic timescale for the exponential growth of the instability is also a factor of approximately 4 shorter for the trimer supermode soliton than the one of the dimer supermode soliton.   

\section{Conclusions} \label{sec:conclusions}

In this study, we explore the emergence of soliton hopping, characterized by the periodic exchange of solitons between coupled optical resonators. We specifically consider photonic dimer and trimer systems, described by sets of coupled LLEs in one periodic spatial dimension.
Augmenting direct numerical simulations, we compute exact invariant solutions including periodic orbits that underlie the temporally periodic soliton hopping. Using numerical continuation tools, we elucidate their bifurcations off stationary soliton equilibrium solutions and characterize the stability of solution branches. 
The soliton hopping solution branches and the Hopf bifurcations creating them are linked to the observed spatio-temporal dynamics close to the bifurcation as captured by the PSD often studied in experiments. This allows interpreting instabilities, bifurcations, and fully nonlinear solution branches in terms of interactions between waves obeying their Nonlinear Dispersion Relation (NDR).

For the dimer system, we identify periodic orbits underlying the soliton hopping dynamics. The branch of periodic orbits bifurcates from a stationary soliton branch in a subcritical Hopf bifurcation. The stationary soliton branch and thus also the periodic orbit branch are initially unstable with respect to a time-periodic `breathing' instability. A stable periodic orbit underlying the robust soliton hopping dynamics observed in the laser scans, is reached after additional saddle-node bifurcations. Considering the dimer problem in the supermode basis, in which coupling between the two supermodes is mediated by the nonlinearity only, allows us to interpret the Hopf bifurcation using wave interactions: The stationary soliton is predominantly confined to the asymmetric supermode, while a neutral eigenvector associated with the Hopf bifurcation has solitonic contributions in the symmetric supermode. The frequency detuning between the two supermode solitons leads to temporal beating and thus the time-periodic nature of the emerging branch of periodic orbits.  

In the trimer system, we also identify a branch of periodic orbits underlying soliton hopping. This soliton hopping branch bifurcates supercritically via a Hopf bifurcation from an equilibrium solution branch that represents a stationary soliton, mostly residing in the S2 supermode. The neutral eigenvectors of the supermode soliton spectrum associated with the Hopf bifurcation display solitonic structures predominantly in the S1 and S3 supermodes. The beating of these solitonic structures with the S2 supermode soliton, which has a frequency offset, yield the time-periodicity of soliton hopping and thereby provides a physical interpretation of the mechanisms associated with the Hopf bifurcation. At the Hopf bifurcation, the stationary soliton branch is unstable so that the emerging periodic orbit is also unstable. The soliton hopping solution branch however stabilizes in an additional saddle-node bifurcation where the stable periodic orbit solution observed in the laser scans is reached. 

Further insights into the Hopf bifurcations creating periodic orbit branches are gained by modifying the forcing protocol from the experimentally accessible single-resonator forcing to a forcing of one supermode only. The dynamics observed in simulated laser scans of the supermode-forced dimers and trimers mimics that observed for the single-resonator forced counterparts and the bifurcations of periodic orbit branches underlying soliton hopping are robust when the forcing protocol is modified. While the characteristics of the Hopf bifurcations creating soliton hopping branches are not modified, the supermode forcing allows the stationary supermode soliton to be strictly confined to the forced supermode without any residual Dispersive Wave (DW) contributions in the othe supermodes, as is the case for the single-resonator forcing. As a consequence, these pure supermode soliton solutions of resonator chains including dimers and trimers can be mapped to well-studied soliton solutions of a single LLE. Moreover, instabilities of the pure supermode solitons can be clearly classified into those constrained within the forced supermode and thus inherited from the single LLE describing a single resonator; and those involving the other supermodes and thus being emergent phenomena due to the coupling between multiple resonators. 

For the supermode-forcing, we obtain periodic orbit solutions describing soliton hopping in dimer and trimer systems that bifurcate from the foliated snaking branches of the supermode soliton solutions, well understood for the single LLE soliton solutions. Both Hopf bifurcations have associated neutrally stable eigenvectors that overlap with the unforced supermodes, thus clearly indicating the soliton hopping phenomenon to be an emergent phenomenon arising from the coupling between different resonators induced by nonlinearity. The dimer soliton hopping branch bifurcates subcritically from the supermode soliton which is stable within the forced supermode, whereas the trimer soliton hopping branch bifurcates supercritically from the supermode soliton branch which is unstable within the supermode.  

The emergence of time-periodic behavior via a Hopf bifurcation is widely observed in many nonlinear pattern-forming systems. However, the considered scenario is often that of a standard supercritical Hopf bifurcation off a stable equilibrium branch, in which the system breaks continuous time-translation symmetry. Below a critical value of the control parameter, the equilibrium is dynamically stable. At the critical value of the Hopf bifurcation, the equilibrium becomes unstable, and a stable limit cycle emerges. In this standard scenario, oscillatory behaviour is expected to commence at the precise critical value of the control parameter with smoothly increasing oscillation amplitude when the control parameter is moved beyond its critical value. There is no coexistence of multiple attracting solutions, no hysteresis, and the emerging branch of stable periodic orbits can often be well described by weakly-nonlinear methods based on small-amplitude approximations close to the bifurcation point. In contrast to this standard scenario,  periodic orbit branches underlying soliton hopping in coupled resontor systems emerge in Hopf bifurcations off unstable equilibrium branches. Moreover, for the dimer, the Hopf bifurcation is also subcritical. The stable periodic orbits underlying robust soliton hopping are only connected to the initial Hopf bifurcation via additional saddle-node bifurcations that stabilize the solutions. As a consequence, the system is expected to show hysteresis, and the observed dynamics can depend not only on the value of all control parameters but the chosen path in parameter space. This is for example evidenced by the rate dependence of the simulated laser scans representing typical experimental measurements: When the detuning parameter $\zeta$ of the dimer is increased at a rate of 0.1 per unit time (see Fig.~\ref{fig:bifhop_5}), the dynamics first approaches the stationary soliton and then reaches the stable soliton hopping periodic orbit at $\zeta=3$, while at slower detuning rates, the instability of the stationary soliton at $\zeta=-0.314$ causes the decay towards the continuous wave background and soliton hopping is not achieved.  

Our analysis shows that in coupled optical resonators, stable periodic orbit branches supporting robust soliton hopping are not directly connected to the Hopf bifurcation creating the periodic orbit branch, but indirectly via additional bifurcations. A more complete understanding of the dynamics of the coupled resonator system thus requires a fully nonlinear bifurcation analysis that traces the entire branch of periodic orbit solutions and determines not only the range of existence but also the range of stability of such solutions. Characterizing the multiple invariant solutions located in the system's state space and their bifurcations as control parameters are varied, moreover helps rationalizing paths in parameter space to reliably reach a certain dynamical response.


\begin{thebibliography}{67}%
\makeatletter
\providecommand \@ifxundefined [1]{%
 \@ifx{#1\undefined}
}%
\providecommand \@ifnum [1]{%
 \ifnum #1\expandafter \@firstoftwo
 \else \expandafter \@secondoftwo
 \fi
}%
\providecommand \@ifx [1]{%
 \ifx #1\expandafter \@firstoftwo
 \else \expandafter \@secondoftwo
 \fi
}%
\providecommand \natexlab [1]{#1}%
\providecommand \enquote  [1]{``#1''}%
\providecommand \bibnamefont  [1]{#1}%
\providecommand \bibfnamefont [1]{#1}%
\providecommand \citenamefont [1]{#1}%
\providecommand \href@noop [0]{\@secondoftwo}%
\providecommand \href [0]{\begingroup \@sanitize@url \@href}%
\providecommand \@href[1]{\@@startlink{#1}\@@href}%
\providecommand \@@href[1]{\endgroup#1\@@endlink}%
\providecommand \@sanitize@url [0]{\catcode `\\12\catcode `\$12\catcode
  `\&12\catcode `\#12\catcode `\^12\catcode `\_12\catcode `\%12\relax}%
\providecommand \@@startlink[1]{}%
\providecommand \@@endlink[0]{}%
\providecommand \url  [0]{\begingroup\@sanitize@url \@url }%
\providecommand \@url [1]{\endgroup\@href {#1}{\urlprefix }}%
\providecommand \urlprefix  [0]{URL }%
\providecommand \Eprint [0]{\href }%
\providecommand \doibase [0]{https://doi.org/}%
\providecommand \selectlanguage [0]{\@gobble}%
\providecommand \bibinfo  [0]{\@secondoftwo}%
\providecommand \bibfield  [0]{\@secondoftwo}%
\providecommand \translation [1]{[#1]}%
\providecommand \BibitemOpen [0]{}%
\providecommand \bibitemStop [0]{}%
\providecommand \bibitemNoStop [0]{.\EOS\space}%
\providecommand \EOS [0]{\spacefactor3000\relax}%
\providecommand \BibitemShut  [1]{\csname bibitem#1\endcsname}%
\let\auto@bib@innerbib\@empty
\bibitem [{\citenamefont {Knobloch}(2015)}]{Knobloch2015Spatial}%
  \BibitemOpen
  \bibfield  {author} {\bibinfo {author} {\bibfnamefont {E.}~\bibnamefont
  {Knobloch}},\ }\bibfield  {title} {\bibinfo {title} {{Spatial localization in
  dissipative systems}},\ }\href
  {https://doi.org/10.1146/annurev-conmatphys-031214-014514} {\bibfield
  {journal} {\bibinfo  {journal} {Annual Review of Condensed Matter Physics}\
  }\textbf {\bibinfo {volume} {6}},\ \bibinfo {pages} {325} (\bibinfo {year}
  {2015})}\BibitemShut {NoStop}%
\bibitem [{\citenamefont {El~Koussaifi}\ \emph {et~al.}(2018)\citenamefont
  {El~Koussaifi}, \citenamefont {Tikan}, \citenamefont {Toffoli}, \citenamefont
  {Randoux}, \citenamefont {Suret},\ and\ \citenamefont
  {Onorato}}]{elkoussaifi2018SpontaneousEmergenceRoguea}%
  \BibitemOpen
  \bibfield  {author} {\bibinfo {author} {\bibfnamefont {R.}~\bibnamefont
  {El~Koussaifi}}, \bibinfo {author} {\bibfnamefont {A.}~\bibnamefont {Tikan}},
  \bibinfo {author} {\bibfnamefont {A.}~\bibnamefont {Toffoli}}, \bibinfo
  {author} {\bibfnamefont {S.}~\bibnamefont {Randoux}}, \bibinfo {author}
  {\bibfnamefont {P.}~\bibnamefont {Suret}},\ and\ \bibinfo {author}
  {\bibfnamefont {M.}~\bibnamefont {Onorato}},\ }\bibfield  {title} {\bibinfo
  {title} {Spontaneous emergence of rogue waves in partially coherent waves:
  {{A}} quantitative experimental comparison between hydrodynamics and
  optics},\ }\href {https://doi.org/10.1103/PhysRevE.97.012208} {\bibfield
  {journal} {\bibinfo  {journal} {Phys. Rev. E}\ }\textbf {\bibinfo {volume}
  {97}},\ \bibinfo {pages} {012208} (\bibinfo {year} {2018})}\BibitemShut
  {NoStop}%
\bibitem [{\citenamefont {Lugiato}\ and\ \citenamefont
  {Lefever}(1987)}]{lugiato1987SpatialDissipativeStructures}%
  \BibitemOpen
  \bibfield  {author} {\bibinfo {author} {\bibfnamefont {L.~A.}\ \bibnamefont
  {Lugiato}}\ and\ \bibinfo {author} {\bibfnamefont {R.}~\bibnamefont
  {Lefever}},\ }\bibfield  {title} {\bibinfo {title} {Spatial {{Dissipative
  Structures}} in {{Passive Optical Systems}}},\ }\href
  {https://doi.org/10.1103/PhysRevLett.58.2209} {\bibfield  {journal} {\bibinfo
   {journal} {Phys. Rev. Lett.}\ }\textbf {\bibinfo {volume} {58}},\ \bibinfo
  {pages} {2209} (\bibinfo {year} {1987})}\BibitemShut {NoStop}%
\bibitem [{\citenamefont {Chembo}\ and\ \citenamefont
  {Yu}(2010)}]{Chembo2010Modal}%
  \BibitemOpen
  \bibfield  {author} {\bibinfo {author} {\bibfnamefont {Y.~K.}\ \bibnamefont
  {Chembo}}\ and\ \bibinfo {author} {\bibfnamefont {N.}~\bibnamefont {Yu}},\
  }\bibfield  {title} {\bibinfo {title} {{Modal expansion approach to
  optical-frequency-comb generation with monolithic whispering-gallery-mode
  resonators}},\ }\href {https://doi.org/10.1103/PhysRevA.82.033801} {\bibfield
   {journal} {\bibinfo  {journal} {Physical Review A - Atomic, Molecular, and
  Optical Physics}\ }\textbf {\bibinfo {volume} {82}},\ \bibinfo {pages} {1}
  (\bibinfo {year} {2010})}\BibitemShut {NoStop}%
\bibitem [{\citenamefont {Chembo}\ and\ \citenamefont
  {Menyuk}(2013)}]{Chembo2013Spatiotemporal}%
  \BibitemOpen
  \bibfield  {author} {\bibinfo {author} {\bibfnamefont {Y.~K.}\ \bibnamefont
  {Chembo}}\ and\ \bibinfo {author} {\bibfnamefont {C.~R.}\ \bibnamefont
  {Menyuk}},\ }\bibfield  {title} {\bibinfo {title} {{Spatiotemporal
  Lugiato-Lefever formalism for Kerr-comb generation in whispering-gallery-mode
  resonators}},\ }\href {https://doi.org/10.1103/PhysRevA.87.053852} {\bibfield
   {journal} {\bibinfo  {journal} {Physical Review A - Atomic, Molecular, and
  Optical Physics}\ }\textbf {\bibinfo {volume} {87}},\ \bibinfo {pages} {1}
  (\bibinfo {year} {2013})}\BibitemShut {NoStop}%
\bibitem [{\citenamefont {Godey}\ \emph {et~al.}(2014)\citenamefont {Godey},
  \citenamefont {Balakireva}, \citenamefont {Coillet},\ and\ \citenamefont
  {Chembo}}]{Godey2014Stability}%
  \BibitemOpen
  \bibfield  {author} {\bibinfo {author} {\bibfnamefont {C.}~\bibnamefont
  {Godey}}, \bibinfo {author} {\bibfnamefont {I.~V.}\ \bibnamefont
  {Balakireva}}, \bibinfo {author} {\bibfnamefont {A.}~\bibnamefont
  {Coillet}},\ and\ \bibinfo {author} {\bibfnamefont {Y.~K.}\ \bibnamefont
  {Chembo}},\ }\bibfield  {title} {\bibinfo {title} {{Stability analysis of the
  spatiotemporal Lugiato-Lefever model for Kerr optical frequency combs in the
  anomalous and normal dispersion regimes}},\ }\href
  {https://doi.org/10.1103/PhysRevA.89.063814} {\bibfield  {journal} {\bibinfo
  {journal} {Physical Review A}\ }\textbf {\bibinfo {volume} {89}},\ \bibinfo
  {pages} {063814} (\bibinfo {year} {2014})}\BibitemShut {NoStop}%
\bibitem [{\citenamefont {Lugiato}\ \emph {et~al.}(2018)\citenamefont
  {Lugiato}, \citenamefont {Prati}, \citenamefont {Gorodetsky},\ and\
  \citenamefont {Kippenberg}}]{lugiato2018LugiatoLefeverEquation}%
  \BibitemOpen
  \bibfield  {author} {\bibinfo {author} {\bibfnamefont {L.~A.}\ \bibnamefont
  {Lugiato}}, \bibinfo {author} {\bibfnamefont {F.}~\bibnamefont {Prati}},
  \bibinfo {author} {\bibfnamefont {M.~L.}\ \bibnamefont {Gorodetsky}},\ and\
  \bibinfo {author} {\bibfnamefont {T.~J.}\ \bibnamefont {Kippenberg}},\
  }\bibfield  {title} {\bibinfo {title} {From the
  lugiato{\textendash}{{Lefever}} equation to microresonator-based soliton kerr
  frequency combs},\ }\bibfield  {journal} {\bibinfo  {journal} {Philosophical
  Transactions of the Royal Society A}\ }\textbf {\bibinfo {volume} {376}},\
  \href {https://doi.org/10.1098/rsta.2018.0113} {10.1098/rsta.2018.0113}
  (\bibinfo {year} {2018})\BibitemShut {NoStop}%
\bibitem [{\citenamefont {Kippenberg}\ \emph {et~al.}(2018)\citenamefont
  {Kippenberg}, \citenamefont {Gaeta}, \citenamefont {Lipson},\ and\
  \citenamefont {Gorodetsky}}]{kippenberg2018DissipativeKerrSolitons}%
  \BibitemOpen
  \bibfield  {author} {\bibinfo {author} {\bibfnamefont {T.~J.}\ \bibnamefont
  {Kippenberg}}, \bibinfo {author} {\bibfnamefont {A.~L.}\ \bibnamefont
  {Gaeta}}, \bibinfo {author} {\bibfnamefont {M.}~\bibnamefont {Lipson}},\ and\
  \bibinfo {author} {\bibfnamefont {M.~L.}\ \bibnamefont {Gorodetsky}},\
  }\bibfield  {title} {\bibinfo {title} {Dissipative {{Kerr}} solitons in
  optical microresonators},\ }\href {https://doi.org/10.1126/science.aan8083}
  {\bibfield  {journal} {\bibinfo  {journal} {Science}\ }\textbf {\bibinfo
  {volume} {361}},\ \bibinfo {pages} {eaan8083} (\bibinfo {year}
  {2018})}\BibitemShut {NoStop}%
\bibitem [{\citenamefont {Pasquazi}\ \emph {et~al.}(2018)\citenamefont
  {Pasquazi}, \citenamefont {Peccianti}, \citenamefont {Razzari}, \citenamefont
  {Moss}, \citenamefont {Coen}, \citenamefont {Erkintalo}, \citenamefont
  {Chembo}, \citenamefont {Hansson}, \citenamefont {Wabnitz}, \citenamefont
  {Del'Haye}, \citenamefont {Xue}, \citenamefont {Weiner},\ and\ \citenamefont
  {Morandotti}}]{pasquazi2018MicrocombsNovelGeneration}%
  \BibitemOpen
  \bibfield  {author} {\bibinfo {author} {\bibfnamefont {A.}~\bibnamefont
  {Pasquazi}}, \bibinfo {author} {\bibfnamefont {M.}~\bibnamefont {Peccianti}},
  \bibinfo {author} {\bibfnamefont {L.}~\bibnamefont {Razzari}}, \bibinfo
  {author} {\bibfnamefont {D.~J.}\ \bibnamefont {Moss}}, \bibinfo {author}
  {\bibfnamefont {S.}~\bibnamefont {Coen}}, \bibinfo {author} {\bibfnamefont
  {M.}~\bibnamefont {Erkintalo}}, \bibinfo {author} {\bibfnamefont {Y.~K.}\
  \bibnamefont {Chembo}}, \bibinfo {author} {\bibfnamefont {T.}~\bibnamefont
  {Hansson}}, \bibinfo {author} {\bibfnamefont {S.}~\bibnamefont {Wabnitz}},
  \bibinfo {author} {\bibfnamefont {P.}~\bibnamefont {Del'Haye}}, \bibinfo
  {author} {\bibfnamefont {X.}~\bibnamefont {Xue}}, \bibinfo {author}
  {\bibfnamefont {A.~M.}\ \bibnamefont {Weiner}},\ and\ \bibinfo {author}
  {\bibfnamefont {R.}~\bibnamefont {Morandotti}},\ }\bibfield  {title}
  {\bibinfo {title} {Micro-combs: {{A}} novel generation of optical sources},\
  }\href {https://doi.org/10.1016/j.physrep.2017.08.004} {\bibfield  {journal}
  {\bibinfo  {journal} {Physics Reports}\ }\textbf {\bibinfo {volume} {729}},\
  \bibinfo {pages} {1} (\bibinfo {year} {2018})}\BibitemShut {NoStop}%
\bibitem [{\citenamefont {Herr}\ \emph {et~al.}(2014)\citenamefont {Herr},
  \citenamefont {Brasch}, \citenamefont {Jost}, \citenamefont {Wang},
  \citenamefont {Kondratiev}, \citenamefont {Gorodetsky},\ and\ \citenamefont
  {Kippenberg}}]{herr2014temporal}%
  \BibitemOpen
  \bibfield  {author} {\bibinfo {author} {\bibfnamefont {T.}~\bibnamefont
  {Herr}}, \bibinfo {author} {\bibfnamefont {V.}~\bibnamefont {Brasch}},
  \bibinfo {author} {\bibfnamefont {J.~D.}\ \bibnamefont {Jost}}, \bibinfo
  {author} {\bibfnamefont {C.~Y.}\ \bibnamefont {Wang}}, \bibinfo {author}
  {\bibfnamefont {N.~M.}\ \bibnamefont {Kondratiev}}, \bibinfo {author}
  {\bibfnamefont {M.~L.}\ \bibnamefont {Gorodetsky}},\ and\ \bibinfo {author}
  {\bibfnamefont {T.~J.}\ \bibnamefont {Kippenberg}},\ }\bibfield  {title}
  {\bibinfo {title} {Temporal solitons in optical microresonators},\
  }\href@noop {} {\bibfield  {journal} {\bibinfo  {journal} {Nature Photonics}\
  }\textbf {\bibinfo {volume} {8}},\ \bibinfo {pages} {145} (\bibinfo {year}
  {2014})}\BibitemShut {NoStop}%
\bibitem [{\citenamefont {Sun}\ \emph {et~al.}(2023)\citenamefont {Sun},
  \citenamefont {Wu}, \citenamefont {Tan}, \citenamefont {Xu}, \citenamefont
  {Li}, \citenamefont {Morandotti}, \citenamefont {Mitchell},\ and\
  \citenamefont {Moss}}]{sun2023ApplicationsOpticalMicrocombs}%
  \BibitemOpen
  \bibfield  {author} {\bibinfo {author} {\bibfnamefont {Y.}~\bibnamefont
  {Sun}}, \bibinfo {author} {\bibfnamefont {J.}~\bibnamefont {Wu}}, \bibinfo
  {author} {\bibfnamefont {M.}~\bibnamefont {Tan}}, \bibinfo {author}
  {\bibfnamefont {X.}~\bibnamefont {Xu}}, \bibinfo {author} {\bibfnamefont
  {Y.}~\bibnamefont {Li}}, \bibinfo {author} {\bibfnamefont {R.}~\bibnamefont
  {Morandotti}}, \bibinfo {author} {\bibfnamefont {A.}~\bibnamefont
  {Mitchell}},\ and\ \bibinfo {author} {\bibfnamefont {D.~J.}\ \bibnamefont
  {Moss}},\ }\bibfield  {title} {\bibinfo {title} {Applications of optical
  microcombs},\ }\href {https://doi.org/10.1364/AOP.470264} {\bibfield
  {journal} {\bibinfo  {journal} {Adv. Opt. Photon.}\ }\textbf {\bibinfo
  {volume} {15}},\ \bibinfo {pages} {86} (\bibinfo {year} {2023})}\BibitemShut
  {NoStop}%
\bibitem [{\citenamefont {Feldmann}\ \emph {et~al.}(2021)\citenamefont
  {Feldmann}, \citenamefont {Youngblood}, \citenamefont {Karpov}, \citenamefont
  {Gehring}, \citenamefont {Li}, \citenamefont {Stappers}, \citenamefont
  {Le~Gallo}, \citenamefont {Fu}, \citenamefont {Lukashchuk}, \citenamefont
  {Raja}, \citenamefont {Liu}, \citenamefont {Wright}, \citenamefont
  {Sebastian}, \citenamefont {Kippenberg}, \citenamefont {Pernice},\ and\
  \citenamefont {Bhaskaran}}]{Feldmann2021ParallelCore}%
  \BibitemOpen
  \bibfield  {author} {\bibinfo {author} {\bibfnamefont {J.}~\bibnamefont
  {Feldmann}}, \bibinfo {author} {\bibfnamefont {N.}~\bibnamefont
  {Youngblood}}, \bibinfo {author} {\bibfnamefont {M.}~\bibnamefont {Karpov}},
  \bibinfo {author} {\bibfnamefont {H.}~\bibnamefont {Gehring}}, \bibinfo
  {author} {\bibfnamefont {X.}~\bibnamefont {Li}}, \bibinfo {author}
  {\bibfnamefont {M.}~\bibnamefont {Stappers}}, \bibinfo {author}
  {\bibfnamefont {M.}~\bibnamefont {Le~Gallo}}, \bibinfo {author}
  {\bibfnamefont {X.}~\bibnamefont {Fu}}, \bibinfo {author} {\bibfnamefont
  {A.}~\bibnamefont {Lukashchuk}}, \bibinfo {author} {\bibfnamefont {A.~S.}\
  \bibnamefont {Raja}}, \bibinfo {author} {\bibfnamefont {J.}~\bibnamefont
  {Liu}}, \bibinfo {author} {\bibfnamefont {C.~D.}\ \bibnamefont {Wright}},
  \bibinfo {author} {\bibfnamefont {A.}~\bibnamefont {Sebastian}}, \bibinfo
  {author} {\bibfnamefont {T.~J.}\ \bibnamefont {Kippenberg}}, \bibinfo
  {author} {\bibfnamefont {W.~H.}\ \bibnamefont {Pernice}},\ and\ \bibinfo
  {author} {\bibfnamefont {H.}~\bibnamefont {Bhaskaran}},\ }\bibfield  {title}
  {\bibinfo {title} {{Parallel convolutional processing using an integrated
  photonic tensor core}},\ }\href {https://doi.org/10.1038/s41586-020-03070-1}
  {\bibfield  {journal} {\bibinfo  {journal} {Nature}\ }\textbf {\bibinfo
  {volume} {589}},\ \bibinfo {pages} {52} (\bibinfo {year} {2021})}\BibitemShut
  {NoStop}%
\bibitem [{\citenamefont {Riemensberger}\ \emph {et~al.}(2020)\citenamefont
  {Riemensberger}, \citenamefont {Lukashchuk}, \citenamefont {Karpov},
  \citenamefont {Weng}, \citenamefont {Lucas}, \citenamefont {Liu},\ and\
  \citenamefont {Kippenberg}}]{riemensberger2020massively}%
  \BibitemOpen
  \bibfield  {author} {\bibinfo {author} {\bibfnamefont {J.}~\bibnamefont
  {Riemensberger}}, \bibinfo {author} {\bibfnamefont {A.}~\bibnamefont
  {Lukashchuk}}, \bibinfo {author} {\bibfnamefont {M.}~\bibnamefont {Karpov}},
  \bibinfo {author} {\bibfnamefont {W.}~\bibnamefont {Weng}}, \bibinfo {author}
  {\bibfnamefont {E.}~\bibnamefont {Lucas}}, \bibinfo {author} {\bibfnamefont
  {J.}~\bibnamefont {Liu}},\ and\ \bibinfo {author} {\bibfnamefont {T.~J.}\
  \bibnamefont {Kippenberg}},\ }\bibfield  {title} {\bibinfo {title} {Massively
  parallel coherent laser ranging using a soliton microcomb},\ }\href@noop {}
  {\bibfield  {journal} {\bibinfo  {journal} {Nature}\ }\textbf {\bibinfo
  {volume} {581}},\ \bibinfo {pages} {164} (\bibinfo {year}
  {2020})}\BibitemShut {NoStop}%
\bibitem [{\citenamefont {Lukashchuk}\ \emph {et~al.}(2021)\citenamefont
  {Lukashchuk}, \citenamefont {Riemensberger}, \citenamefont {Tusnin},
  \citenamefont {Liu},\ and\ \citenamefont
  {Kippenberg}}]{lukashchuk2021chaotic}%
  \BibitemOpen
  \bibfield  {author} {\bibinfo {author} {\bibfnamefont {A.}~\bibnamefont
  {Lukashchuk}}, \bibinfo {author} {\bibfnamefont {J.}~\bibnamefont
  {Riemensberger}}, \bibinfo {author} {\bibfnamefont {A.}~\bibnamefont
  {Tusnin}}, \bibinfo {author} {\bibfnamefont {J.}~\bibnamefont {Liu}},\ and\
  \bibinfo {author} {\bibfnamefont {T.}~\bibnamefont {Kippenberg}},\ }\bibfield
   {title} {\bibinfo {title} {Chaotic micro-comb based parallel ranging},\
  }\href@noop {} {\bibfield  {journal} {\bibinfo  {journal} {arXiv preprint
  arXiv:2112.10241}\ } (\bibinfo {year} {2021})}\BibitemShut {NoStop}%
\bibitem [{\citenamefont {Marin-Palomo}\ \emph {et~al.}(2017)\citenamefont
  {Marin-Palomo}, \citenamefont {Kemal}, \citenamefont {Karpov}, \citenamefont
  {Kordts}, \citenamefont {Pfeifle}, \citenamefont {Pfeiffer}, \citenamefont
  {Trocha}, \citenamefont {Wolf}, \citenamefont {Brasch}, \citenamefont
  {Anderson}, \citenamefont {Rosenberger}, \citenamefont {Vijayan},
  \citenamefont {Freude}, \citenamefont {Kippenberg},\ and\ \citenamefont
  {Koos}}]{Marin-Palomo2017}%
  \BibitemOpen
  \bibfield  {author} {\bibinfo {author} {\bibfnamefont {P.}~\bibnamefont
  {Marin-Palomo}}, \bibinfo {author} {\bibfnamefont {J.~N.}\ \bibnamefont
  {Kemal}}, \bibinfo {author} {\bibfnamefont {M.}~\bibnamefont {Karpov}},
  \bibinfo {author} {\bibfnamefont {A.}~\bibnamefont {Kordts}}, \bibinfo
  {author} {\bibfnamefont {J.}~\bibnamefont {Pfeifle}}, \bibinfo {author}
  {\bibfnamefont {M.~H.}\ \bibnamefont {Pfeiffer}}, \bibinfo {author}
  {\bibfnamefont {P.}~\bibnamefont {Trocha}}, \bibinfo {author} {\bibfnamefont
  {S.}~\bibnamefont {Wolf}}, \bibinfo {author} {\bibfnamefont {V.}~\bibnamefont
  {Brasch}}, \bibinfo {author} {\bibfnamefont {M.~H.}\ \bibnamefont
  {Anderson}}, \bibinfo {author} {\bibfnamefont {R.}~\bibnamefont
  {Rosenberger}}, \bibinfo {author} {\bibfnamefont {K.}~\bibnamefont
  {Vijayan}}, \bibinfo {author} {\bibfnamefont {W.}~\bibnamefont {Freude}},
  \bibinfo {author} {\bibfnamefont {T.~J.}\ \bibnamefont {Kippenberg}},\ and\
  \bibinfo {author} {\bibfnamefont {C.}~\bibnamefont {Koos}},\ }\bibfield
  {title} {\bibinfo {title} {{Microresonator-based solitons for massively
  parallel coherent optical communications}},\ }\href
  {https://doi.org/10.1038/nature22387} {\bibfield  {journal} {\bibinfo
  {journal} {Nature}\ }\textbf {\bibinfo {volume} {546}},\ \bibinfo {pages}
  {274} (\bibinfo {year} {2017})}\BibitemShut {NoStop}%
\bibitem [{\citenamefont {Liu}\ \emph {et~al.}(2021)\citenamefont {Liu},
  \citenamefont {Huang}, \citenamefont {Wang}, \citenamefont {He},
  \citenamefont {Raja}, \citenamefont {Liu}, \citenamefont {Engelsen},\ and\
  \citenamefont {Kippenberg}}]{liu2021high}%
  \BibitemOpen
  \bibfield  {author} {\bibinfo {author} {\bibfnamefont {J.}~\bibnamefont
  {Liu}}, \bibinfo {author} {\bibfnamefont {G.}~\bibnamefont {Huang}}, \bibinfo
  {author} {\bibfnamefont {R.~N.}\ \bibnamefont {Wang}}, \bibinfo {author}
  {\bibfnamefont {J.}~\bibnamefont {He}}, \bibinfo {author} {\bibfnamefont
  {A.~S.}\ \bibnamefont {Raja}}, \bibinfo {author} {\bibfnamefont
  {T.}~\bibnamefont {Liu}}, \bibinfo {author} {\bibfnamefont {N.~J.}\
  \bibnamefont {Engelsen}},\ and\ \bibinfo {author} {\bibfnamefont {T.~J.}\
  \bibnamefont {Kippenberg}},\ }\bibfield  {title} {\bibinfo {title}
  {High-yield, wafer-scale fabrication of ultralow-loss, dispersion-engineered
  silicon nitride photonic circuits},\ }\href@noop {} {\bibfield  {journal}
  {\bibinfo  {journal} {Nature communications}\ }\textbf {\bibinfo {volume}
  {12}},\ \bibinfo {pages} {1} (\bibinfo {year} {2021})}\BibitemShut {NoStop}%
\bibitem [{\citenamefont {Tikan}\ \emph {et~al.}(2021)\citenamefont {Tikan},
  \citenamefont {Riemensberger}, \citenamefont {Komagata}, \citenamefont
  {H{\"o}nl}, \citenamefont {Churaev}, \citenamefont {Skehan}, \citenamefont
  {Guo}, \citenamefont {Wang}, \citenamefont {Liu}, \citenamefont {Seidler},\
  and\ \citenamefont {Kippenberg}}]{tikan2021emergent}%
  \BibitemOpen
  \bibfield  {author} {\bibinfo {author} {\bibfnamefont {A.}~\bibnamefont
  {Tikan}}, \bibinfo {author} {\bibfnamefont {J.}~\bibnamefont
  {Riemensberger}}, \bibinfo {author} {\bibfnamefont {K.}~\bibnamefont
  {Komagata}}, \bibinfo {author} {\bibfnamefont {S.}~\bibnamefont {H{\"o}nl}},
  \bibinfo {author} {\bibfnamefont {M.}~\bibnamefont {Churaev}}, \bibinfo
  {author} {\bibfnamefont {C.}~\bibnamefont {Skehan}}, \bibinfo {author}
  {\bibfnamefont {H.}~\bibnamefont {Guo}}, \bibinfo {author} {\bibfnamefont
  {R.~N.}\ \bibnamefont {Wang}}, \bibinfo {author} {\bibfnamefont
  {J.}~\bibnamefont {Liu}}, \bibinfo {author} {\bibfnamefont {P.}~\bibnamefont
  {Seidler}},\ and\ \bibinfo {author} {\bibfnamefont {T.~J.}\ \bibnamefont
  {Kippenberg}},\ }\bibfield  {title} {\bibinfo {title} {Emergent nonlinear
  phenomena in a driven dissipative photonic dimer},\ }\href
  {https://doi.org/10.1038/s41567-020-01159-y} {\bibfield  {journal} {\bibinfo
  {journal} {Nature Physics}\ }\textbf {\bibinfo {volume} {17}},\ \bibinfo
  {pages} {604} (\bibinfo {year} {2021})}\BibitemShut {NoStop}%
\bibitem [{\citenamefont {Tikan}\ \emph
  {et~al.}(2022{\natexlab{a}})\citenamefont {Tikan}, \citenamefont {Tusnin},
  \citenamefont {Riemensberger}, \citenamefont {Churaev}, \citenamefont {Ji},
  \citenamefont {Komagata}, \citenamefont {Wang}, \citenamefont {Liu},\ and\
  \citenamefont {Kippenberg}}]{tikan2022protected}%
  \BibitemOpen
  \bibfield  {author} {\bibinfo {author} {\bibfnamefont {A.}~\bibnamefont
  {Tikan}}, \bibinfo {author} {\bibfnamefont {A.}~\bibnamefont {Tusnin}},
  \bibinfo {author} {\bibfnamefont {J.}~\bibnamefont {Riemensberger}}, \bibinfo
  {author} {\bibfnamefont {M.}~\bibnamefont {Churaev}}, \bibinfo {author}
  {\bibfnamefont {X.}~\bibnamefont {Ji}}, \bibinfo {author} {\bibfnamefont
  {K.}~\bibnamefont {Komagata}}, \bibinfo {author} {\bibfnamefont {R.~N.}\
  \bibnamefont {Wang}}, \bibinfo {author} {\bibfnamefont {J.}~\bibnamefont
  {Liu}},\ and\ \bibinfo {author} {\bibfnamefont {T.~J.}\ \bibnamefont
  {Kippenberg}},\ }\bibfield  {title} {\bibinfo {title} {{Protected generation
  of dissipative Kerr solitons in supermodes of coupled optical
  microresonators}},\ }\href {https://doi.org/10.1126/sciadv.abm6982}
  {\bibfield  {journal} {\bibinfo  {journal} {Science Advances}\ }\textbf
  {\bibinfo {volume} {8}},\ \bibinfo {pages} {eabm6982} (\bibinfo {year}
  {2022}{\natexlab{a}})}\BibitemShut {NoStop}%
\bibitem [{\citenamefont {Helgason}\ \emph
  {et~al.}(2021{\natexlab{a}})\citenamefont {Helgason}, \citenamefont
  {Arteaga-Sierra}, \citenamefont {Ye}, \citenamefont {Twayana}, \citenamefont
  {Andrekson}, \citenamefont {Karlsson}, \citenamefont {Schr{\"o}der},\ and\
  \citenamefont {Torres-Company}}]{helgason2021dissipative}%
  \BibitemOpen
  \bibfield  {author} {\bibinfo {author} {\bibfnamefont {{\'O}.~B.}\
  \bibnamefont {Helgason}}, \bibinfo {author} {\bibfnamefont {F.~R.}\
  \bibnamefont {Arteaga-Sierra}}, \bibinfo {author} {\bibfnamefont
  {Z.}~\bibnamefont {Ye}}, \bibinfo {author} {\bibfnamefont {K.}~\bibnamefont
  {Twayana}}, \bibinfo {author} {\bibfnamefont {P.~A.}\ \bibnamefont
  {Andrekson}}, \bibinfo {author} {\bibfnamefont {M.}~\bibnamefont {Karlsson}},
  \bibinfo {author} {\bibfnamefont {J.}~\bibnamefont {Schr{\"o}der}},\ and\
  \bibinfo {author} {\bibfnamefont {V.}~\bibnamefont {Torres-Company}},\
  }\bibfield  {title} {\bibinfo {title} {Dissipative solitons in photonic
  molecules},\ }\href@noop {} {\bibfield  {journal} {\bibinfo  {journal}
  {Nature Photonics}\ }\textbf {\bibinfo {volume} {15}},\ \bibinfo {pages}
  {305} (\bibinfo {year} {2021}{\natexlab{a}})}\BibitemShut {NoStop}%
\bibitem [{\citenamefont {Ji}\ \emph {et~al.}(2023)\citenamefont {Ji},
  \citenamefont {Jin}, \citenamefont {Wu}, \citenamefont {Yu}, \citenamefont
  {Yuan}, \citenamefont {Zhang}, \citenamefont {Gao}, \citenamefont {Li},
  \citenamefont {Wang}, \citenamefont {Xiang}, \citenamefont {Guo},
  \citenamefont {Feshali}, \citenamefont {Paniccia}, \citenamefont {Ilchenko},
  \citenamefont {Matsko}, \citenamefont {Bowers},\ and\ \citenamefont
  {Vahala}}]{ji2023EngineeredZerodispersionMicrocombsa}%
  \BibitemOpen
  \bibfield  {author} {\bibinfo {author} {\bibfnamefont {Q.-X.}\ \bibnamefont
  {Ji}}, \bibinfo {author} {\bibfnamefont {W.}~\bibnamefont {Jin}}, \bibinfo
  {author} {\bibfnamefont {L.}~\bibnamefont {Wu}}, \bibinfo {author}
  {\bibfnamefont {Y.}~\bibnamefont {Yu}}, \bibinfo {author} {\bibfnamefont
  {Z.}~\bibnamefont {Yuan}}, \bibinfo {author} {\bibfnamefont {W.}~\bibnamefont
  {Zhang}}, \bibinfo {author} {\bibfnamefont {M.}~\bibnamefont {Gao}}, \bibinfo
  {author} {\bibfnamefont {B.}~\bibnamefont {Li}}, \bibinfo {author}
  {\bibfnamefont {H.}~\bibnamefont {Wang}}, \bibinfo {author} {\bibfnamefont
  {C.}~\bibnamefont {Xiang}}, \bibinfo {author} {\bibfnamefont
  {J.}~\bibnamefont {Guo}}, \bibinfo {author} {\bibfnamefont {A.}~\bibnamefont
  {Feshali}}, \bibinfo {author} {\bibfnamefont {M.}~\bibnamefont {Paniccia}},
  \bibinfo {author} {\bibfnamefont {V.~S.}\ \bibnamefont {Ilchenko}}, \bibinfo
  {author} {\bibfnamefont {A.~B.}\ \bibnamefont {Matsko}}, \bibinfo {author}
  {\bibfnamefont {J.~E.}\ \bibnamefont {Bowers}},\ and\ \bibinfo {author}
  {\bibfnamefont {K.~J.}\ \bibnamefont {Vahala}},\ }\bibfield  {title}
  {\bibinfo {title} {Engineered zero-dispersion microcombs using {{CMOS-ready}}
  photonics},\ }\href {https://doi.org/10.1364/OPTICA.478710} {\bibfield
  {journal} {\bibinfo  {journal} {Optica}\ }\textbf {\bibinfo {volume} {10}},\
  \bibinfo {pages} {279} (\bibinfo {year} {2023})}\BibitemShut {NoStop}%
\bibitem [{\citenamefont {Yuan}\ \emph {et~al.}(2023)\citenamefont {Yuan},
  \citenamefont {Gao}, \citenamefont {Yu}, \citenamefont {Wang}, \citenamefont
  {Jin}, \citenamefont {Ji}, \citenamefont {Feshali}, \citenamefont {Paniccia},
  \citenamefont {Bowers},\ and\ \citenamefont
  {Vahala}}]{yuan2023SolitonPulsePairs}%
  \BibitemOpen
  \bibfield  {author} {\bibinfo {author} {\bibfnamefont {Z.}~\bibnamefont
  {Yuan}}, \bibinfo {author} {\bibfnamefont {M.}~\bibnamefont {Gao}}, \bibinfo
  {author} {\bibfnamefont {Y.}~\bibnamefont {Yu}}, \bibinfo {author}
  {\bibfnamefont {H.}~\bibnamefont {Wang}}, \bibinfo {author} {\bibfnamefont
  {W.}~\bibnamefont {Jin}}, \bibinfo {author} {\bibfnamefont {Q.-X.}\
  \bibnamefont {Ji}}, \bibinfo {author} {\bibfnamefont {A.}~\bibnamefont
  {Feshali}}, \bibinfo {author} {\bibfnamefont {M.}~\bibnamefont {Paniccia}},
  \bibinfo {author} {\bibfnamefont {J.}~\bibnamefont {Bowers}},\ and\ \bibinfo
  {author} {\bibfnamefont {K.}~\bibnamefont {Vahala}},\ }\bibfield  {title}
  {\bibinfo {title} {Soliton pulse pairs at multiple colours in normal
  dispersion microresonators},\ }\href
  {https://doi.org/10.1038/s41566-023-01257-2} {\bibfield  {journal} {\bibinfo
  {journal} {Nat. Photon.}\ ,\ \bibinfo {pages} {1}} (\bibinfo {year}
  {2023})}\BibitemShut {NoStop}%
\bibitem [{\citenamefont {Helgason}\ \emph {et~al.}(2023)\citenamefont
  {Helgason}, \citenamefont {Girardi}, \citenamefont {Ye}, \citenamefont {Lei},
  \citenamefont {Schr{\"o}der},\ and\ \citenamefont
  {{Torres-Company}}}]{helgason2023SurpassingNonlinearConversion}%
  \BibitemOpen
  \bibfield  {author} {\bibinfo {author} {\bibfnamefont {{\'O}.~B.}\
  \bibnamefont {Helgason}}, \bibinfo {author} {\bibfnamefont {M.}~\bibnamefont
  {Girardi}}, \bibinfo {author} {\bibfnamefont {Z.}~\bibnamefont {Ye}},
  \bibinfo {author} {\bibfnamefont {F.}~\bibnamefont {Lei}}, \bibinfo {author}
  {\bibfnamefont {J.}~\bibnamefont {Schr{\"o}der}},\ and\ \bibinfo {author}
  {\bibfnamefont {V.}~\bibnamefont {{Torres-Company}}},\ }\bibfield  {title}
  {\bibinfo {title} {Surpassing the nonlinear conversion efficiency of soliton
  microcombs},\ }\href {https://doi.org/10.1038/s41566-023-01280-3} {\bibfield
  {journal} {\bibinfo  {journal} {Nat. Photon.}\ ,\ \bibinfo {pages} {1}}
  (\bibinfo {year} {2023})}\BibitemShut {NoStop}%
\bibitem [{\citenamefont {Xue}\ \emph {et~al.}(2015)\citenamefont {Xue},
  \citenamefont {Xuan}, \citenamefont {Wang}, \citenamefont {Liu},
  \citenamefont {Leaird}, \citenamefont {Qi},\ and\ \citenamefont
  {Weiner}}]{xue2015NormaldispersionMicrocombsEnabled}%
  \BibitemOpen
  \bibfield  {author} {\bibinfo {author} {\bibfnamefont {X.}~\bibnamefont
  {Xue}}, \bibinfo {author} {\bibfnamefont {Y.}~\bibnamefont {Xuan}}, \bibinfo
  {author} {\bibfnamefont {P.-H.}\ \bibnamefont {Wang}}, \bibinfo {author}
  {\bibfnamefont {Y.}~\bibnamefont {Liu}}, \bibinfo {author} {\bibfnamefont
  {D.~E.}\ \bibnamefont {Leaird}}, \bibinfo {author} {\bibfnamefont
  {M.}~\bibnamefont {Qi}},\ and\ \bibinfo {author} {\bibfnamefont {A.~M.}\
  \bibnamefont {Weiner}},\ }\bibfield  {title} {\bibinfo {title}
  {Normal-dispersion microcombs enabled by controllable mode interactions},\
  }\href {https://doi.org/10.1002/lpor.201500107} {\bibfield  {journal}
  {\bibinfo  {journal} {Laser \& Photonics Reviews}\ }\textbf {\bibinfo
  {volume} {9}},\ \bibinfo {pages} {L23} (\bibinfo {year} {2015})}\BibitemShut
  {NoStop}%
\bibitem [{\citenamefont {Helgason}\ \emph
  {et~al.}(2021{\natexlab{b}})\citenamefont {Helgason}, \citenamefont
  {{Arteaga-Sierra}}, \citenamefont {Ye}, \citenamefont {Twayana},
  \citenamefont {Andrekson}, \citenamefont {Karlsson}, \citenamefont
  {Schr{\"o}der},\ and\ \citenamefont {{Victor
  Torres-Company}}}]{helgason2021DissipativeSolitonsPhotonic}%
  \BibitemOpen
  \bibfield  {author} {\bibinfo {author} {\bibfnamefont {{\'O}.~B.}\
  \bibnamefont {Helgason}}, \bibinfo {author} {\bibfnamefont {F.~R.}\
  \bibnamefont {{Arteaga-Sierra}}}, \bibinfo {author} {\bibfnamefont
  {Z.}~\bibnamefont {Ye}}, \bibinfo {author} {\bibfnamefont {K.}~\bibnamefont
  {Twayana}}, \bibinfo {author} {\bibfnamefont {P.~A.}\ \bibnamefont
  {Andrekson}}, \bibinfo {author} {\bibfnamefont {M.}~\bibnamefont {Karlsson}},
  \bibinfo {author} {\bibfnamefont {J.}~\bibnamefont {Schr{\"o}der}},\ and\
  \bibinfo {author} {\bibnamefont {{Victor Torres-Company}}},\ }\bibfield
  {title} {\bibinfo {title} {Dissipative solitons in photonic molecules},\
  }\href {https://doi.org/10.1038/s41566-020-00757-9} {\bibfield  {journal}
  {\bibinfo  {journal} {Nat. Photonics}\ }\textbf {\bibinfo {volume} {15}},\
  \bibinfo {pages} {305} (\bibinfo {year} {2021}{\natexlab{b}})}\BibitemShut
  {NoStop}%
\bibitem [{\citenamefont {Pidgayko}\ \emph {et~al.}(2023)\citenamefont
  {Pidgayko}, \citenamefont {Tusnin}, \citenamefont {Riemensberger},
  \citenamefont {Stroganov}, \citenamefont {Tikan},\ and\ \citenamefont
  {Kippenberg}}]{pidgayko2023voltage}%
  \BibitemOpen
  \bibfield  {author} {\bibinfo {author} {\bibfnamefont {D.}~\bibnamefont
  {Pidgayko}}, \bibinfo {author} {\bibfnamefont {A.}~\bibnamefont {Tusnin}},
  \bibinfo {author} {\bibfnamefont {J.}~\bibnamefont {Riemensberger}}, \bibinfo
  {author} {\bibfnamefont {A.}~\bibnamefont {Stroganov}}, \bibinfo {author}
  {\bibfnamefont {A.}~\bibnamefont {Tikan}},\ and\ \bibinfo {author}
  {\bibfnamefont {T.~J.}\ \bibnamefont {Kippenberg}},\ }\bibfield  {title}
  {\bibinfo {title} {Voltage-tunable opo with an alternating dispersion dimer
  integrated on chip},\ }\href@noop {} {\bibfield  {journal} {\bibinfo
  {journal} {arXiv preprint arXiv:2308.01951}\ } (\bibinfo {year}
  {2023})}\BibitemShut {NoStop}%
\bibitem [{\citenamefont {Ghosh}\ \emph {et~al.}(2024)\citenamefont {Ghosh},
  \citenamefont {Pal}, \citenamefont {Hill}, \citenamefont {Campbell},
  \citenamefont {Bi}, \citenamefont {Zhang}, \citenamefont {Alabbadi},
  \citenamefont {Zhang},\ and\ \citenamefont {{Del Haye}}}]{Ghosh2024}%
  \BibitemOpen
  \bibfield  {author} {\bibinfo {author} {\bibfnamefont {A.}~\bibnamefont
  {Ghosh}}, \bibinfo {author} {\bibfnamefont {A.}~\bibnamefont {Pal}}, \bibinfo
  {author} {\bibfnamefont {L.}~\bibnamefont {Hill}}, \bibinfo {author}
  {\bibfnamefont {G.}~\bibnamefont {Campbell}}, \bibinfo {author}
  {\bibfnamefont {T.}~\bibnamefont {Bi}}, \bibinfo {author} {\bibfnamefont
  {Y.}~\bibnamefont {Zhang}}, \bibinfo {author} {\bibfnamefont
  {A.}~\bibnamefont {Alabbadi}}, \bibinfo {author} {\bibfnamefont
  {S.}~\bibnamefont {Zhang}},\ and\ \bibinfo {author} {\bibfnamefont
  {P.}~\bibnamefont {{Del Haye}}},\ }\bibfield  {title} {\bibinfo {title}
  {{Controlled light distribution with coupled microresonator chains via Kerr
  symmetry breaking}},\ }\href {https://doi.org/10.1364/prj.524823} {\bibfield
  {journal} {\bibinfo  {journal} {Photonics Research}\ }\textbf {\bibinfo
  {volume} {12}},\ \bibinfo {pages} {2376} (\bibinfo {year} {2024})},\ \Eprint
  {https://arxiv.org/abs/2402.10673} {arXiv:2402.10673} \BibitemShut {NoStop}%
\bibitem [{\citenamefont {Morichetti}\ \emph {et~al.}(2012)\citenamefont
  {Morichetti}, \citenamefont {Ferrari}, \citenamefont {Canciamilla},\ and\
  \citenamefont {Melloni}}]{Morichetti2012First}%
  \BibitemOpen
  \bibfield  {author} {\bibinfo {author} {\bibfnamefont {F.}~\bibnamefont
  {Morichetti}}, \bibinfo {author} {\bibfnamefont {C.}~\bibnamefont {Ferrari}},
  \bibinfo {author} {\bibfnamefont {A.}~\bibnamefont {Canciamilla}},\ and\
  \bibinfo {author} {\bibfnamefont {A.}~\bibnamefont {Melloni}},\ }\bibfield
  {title} {\bibinfo {title} {{The first decade of coupled resonator optical
  waveguides: Bringing slow light to applications}},\ }\href
  {https://doi.org/10.1002/lpor.201100018} {\bibfield  {journal} {\bibinfo
  {journal} {Laser and Photonics Reviews}\ }\textbf {\bibinfo {volume} {6}},\
  \bibinfo {pages} {74} (\bibinfo {year} {2012})}\BibitemShut {NoStop}%
\bibitem [{\citenamefont {Hopf}(1948)}]{Hopf1948}%
  \BibitemOpen
  \bibfield  {author} {\bibinfo {author} {\bibfnamefont {E.}~\bibnamefont
  {Hopf}},\ }\bibfield  {title} {\bibinfo {title} {A mathematical example
  displaying features of turbulence},\ }\href@noop {} {\bibfield  {journal}
  {\bibinfo  {journal} {Communications on Pure and Applied Mathematics}\
  }\textbf {\bibinfo {volume} {1}},\ \bibinfo {pages} {303} (\bibinfo {year}
  {1948})}\BibitemShut {NoStop}%
\bibitem [{\citenamefont {Nagata}(1990)}]{Nagataf1990}%
  \BibitemOpen
  \bibfield  {author} {\bibinfo {author} {\bibfnamefont {M.}~\bibnamefont
  {Nagata}},\ }\bibfield  {title} {\bibinfo {title} {{Three-dimensional
  finite-amplitude solutions in plane Couette flow: bifurcation from
  infinity}},\ }\href {https://doi.org/10.1017/S0022112090000829} {\bibfield
  {journal} {\bibinfo  {journal} {Journal of Fluid Mechanics}\ }\textbf
  {\bibinfo {volume} {217}},\ \bibinfo {pages} {519} (\bibinfo {year}
  {1990})}\BibitemShut {NoStop}%
\bibitem [{\citenamefont {Kerswell}(2005)}]{Kerswell2005}%
  \BibitemOpen
  \bibfield  {author} {\bibinfo {author} {\bibfnamefont {R.~R.}\ \bibnamefont
  {Kerswell}},\ }\bibfield  {title} {\bibinfo {title} {{Recent progress in
  understanding the transition to turbulence in a pipe}},\ }\href
  {https://doi.org/10.1088/0951-7715/18/6/R01} {\bibfield  {journal} {\bibinfo
  {journal} {Nonlinearity}\ }\textbf {\bibinfo {volume} {18}},\ \bibinfo
  {pages} {R17} (\bibinfo {year} {2005})}\BibitemShut {NoStop}%
\bibitem [{\citenamefont {Gibson}\ \emph {et~al.}(2008)\citenamefont {Gibson},
  \citenamefont {Halcrow},\ and\ \citenamefont {Cvitanović}}]{Gibson2008}%
  \BibitemOpen
  \bibfield  {author} {\bibinfo {author} {\bibfnamefont {J.~F.}\ \bibnamefont
  {Gibson}}, \bibinfo {author} {\bibfnamefont {J.}~\bibnamefont {Halcrow}},\
  and\ \bibinfo {author} {\bibfnamefont {P.}~\bibnamefont {Cvitanović}},\
  }\bibfield  {title} {\bibinfo {title} {Visualizing the geometry of state
  space in plane couette flow},\ }\href
  {https://doi.org/10.1017/S002211200800267X} {\bibfield  {journal} {\bibinfo
  {journal} {Journal of Fluid Mechanics}\ }\textbf {\bibinfo {volume} {611}},\
  \bibinfo {pages} {107} (\bibinfo {year} {2008})}\BibitemShut {NoStop}%
\bibitem [{\citenamefont {Kawahara}\ \emph {et~al.}(2011)\citenamefont
  {Kawahara}, \citenamefont {Uhlmann},\ and\ \citenamefont {{Van
  Veen}}}]{Kawahara2011}%
  \BibitemOpen
  \bibfield  {author} {\bibinfo {author} {\bibfnamefont {G.}~\bibnamefont
  {Kawahara}}, \bibinfo {author} {\bibfnamefont {M.}~\bibnamefont {Uhlmann}},\
  and\ \bibinfo {author} {\bibfnamefont {L.}~\bibnamefont {{Van Veen}}},\
  }\bibfield  {title} {\bibinfo {title} {{The significance of simple invariant
  solutions in turbulent flows}},\ }\href
  {https://doi.org/10.1146/annurev-fluid-120710-101228} {\bibfield  {journal}
  {\bibinfo  {journal} {Annual Review of Fluid Mechanics}\ }\textbf {\bibinfo
  {volume} {44}},\ \bibinfo {pages} {203} (\bibinfo {year} {2011})},\ \Eprint
  {https://arxiv.org/abs/1108.0975} {arXiv:1108.0975} \BibitemShut {NoStop}%
\bibitem [{\citenamefont {Suri}\ \emph {et~al.}(2017)\citenamefont {Suri},
  \citenamefont {Tithof}, \citenamefont {Grigoriev},\ and\ \citenamefont
  {Schatz}}]{Suri2017}%
  \BibitemOpen
  \bibfield  {author} {\bibinfo {author} {\bibfnamefont {B.}~\bibnamefont
  {Suri}}, \bibinfo {author} {\bibfnamefont {J.}~\bibnamefont {Tithof}},
  \bibinfo {author} {\bibfnamefont {R.~O.}\ \bibnamefont {Grigoriev}},\ and\
  \bibinfo {author} {\bibfnamefont {M.~F.}\ \bibnamefont {Schatz}},\ }\bibfield
   {title} {\bibinfo {title} {Forecasting fluid flows using the geometry of
  turbulence},\ }\href {https://doi.org/10.1103/PhysRevLett.118.114501}
  {\bibfield  {journal} {\bibinfo  {journal} {Physical Review Letters}\
  }\textbf {\bibinfo {volume} {118}},\ \bibinfo {pages} {114501} (\bibinfo
  {year} {2017})}\BibitemShut {NoStop}%
\bibitem [{\citenamefont {Graham}\ and\ \citenamefont
  {Floryan}(2021)}]{Graham2021}%
  \BibitemOpen
  \bibfield  {author} {\bibinfo {author} {\bibfnamefont {M.~D.}\ \bibnamefont
  {Graham}}\ and\ \bibinfo {author} {\bibfnamefont {D.}~\bibnamefont
  {Floryan}},\ }\bibfield  {title} {\bibinfo {title} {{Exact Coherent States
  and the Nonlinear Dynamics of Wall-Bounded Turbulent Flows}},\ }\href
  {https://doi.org/10.1146/annurev-fluid-051820-020223} {\bibfield  {journal}
  {\bibinfo  {journal} {Annual Review of Fluid Mechanics}\ }\textbf {\bibinfo
  {volume} {53}},\ \bibinfo {pages} {227} (\bibinfo {year} {2021})}\BibitemShut
  {NoStop}%
\bibitem [{\citenamefont {Reetz}\ \emph {et~al.}(2019)\citenamefont {Reetz},
  \citenamefont {Kreilos},\ and\ \citenamefont {Schneider}}]{Reetz2019a}%
  \BibitemOpen
  \bibfield  {author} {\bibinfo {author} {\bibfnamefont {F.}~\bibnamefont
  {Reetz}}, \bibinfo {author} {\bibfnamefont {T.}~\bibnamefont {Kreilos}},\
  and\ \bibinfo {author} {\bibfnamefont {T.~M.}\ \bibnamefont {Schneider}},\
  }\bibfield  {title} {\bibinfo {title} {Exact invariant solution reveals the
  origin of self-organized oblique turbulent-laminar stripes},\ }\href
  {https://doi.org/10.1038/s41467-019-10208-x} {\bibfield  {journal} {\bibinfo
  {journal} {Nature Communications}\ }\textbf {\bibinfo {volume} {10}},\
  \bibinfo {pages} {2277} (\bibinfo {year} {2019})}\BibitemShut {NoStop}%
\bibitem [{\citenamefont {Crowley}\ \emph {et~al.}(2022)\citenamefont
  {Crowley}, \citenamefont {Pughe-Sanford}, \citenamefont {Toler},
  \citenamefont {Krygier}, \citenamefont {Grigoriev},\ and\ \citenamefont
  {Schatz}}]{Crowley2022}%
  \BibitemOpen
  \bibfield  {author} {\bibinfo {author} {\bibfnamefont {C.~J.}\ \bibnamefont
  {Crowley}}, \bibinfo {author} {\bibfnamefont {J.~L.}\ \bibnamefont
  {Pughe-Sanford}}, \bibinfo {author} {\bibfnamefont {W.}~\bibnamefont
  {Toler}}, \bibinfo {author} {\bibfnamefont {M.~C.}\ \bibnamefont {Krygier}},
  \bibinfo {author} {\bibfnamefont {R.~O.}\ \bibnamefont {Grigoriev}},\ and\
  \bibinfo {author} {\bibfnamefont {M.~F.}\ \bibnamefont {Schatz}},\ }\bibfield
   {title} {\bibinfo {title} {{Turbulence tracks recurrent solutions}},\ }\href
  {https://doi.org/10.1073/pnas.2120665119} {\bibfield  {journal} {\bibinfo
  {journal} {Proceedings of the National Academy of Sciences of the United
  States of America}\ }\textbf {\bibinfo {volume} {119}},\ \bibinfo {pages} {1}
  (\bibinfo {year} {2022})}\BibitemShut {NoStop}%
\bibitem [{\citenamefont {Bodenschatz}\ \emph {et~al.}(2000)\citenamefont
  {Bodenschatz}, \citenamefont {Pesch},\ and\ \citenamefont
  {Ahlers}}]{Bodenschatz2000}%
  \BibitemOpen
  \bibfield  {author} {\bibinfo {author} {\bibfnamefont {E.}~\bibnamefont
  {Bodenschatz}}, \bibinfo {author} {\bibfnamefont {W.}~\bibnamefont {Pesch}},\
  and\ \bibinfo {author} {\bibfnamefont {G.}~\bibnamefont {Ahlers}},\
  }\bibfield  {title} {\bibinfo {title} {Recent developments in
  rayleigh-b{\'e}nard convection},\ }\href
  {https://doi.org/10.1146/annurev.fluid.32.1.709} {\bibfield  {journal}
  {\bibinfo  {journal} {Annu. Rev. Fluid Mech.}\ }\textbf {\bibinfo {volume}
  {32}},\ \bibinfo {pages} {709} (\bibinfo {year} {2000})}\BibitemShut
  {NoStop}%
\bibitem [{\citenamefont {Lappa}(2009)}]{Lappa2009}%
  \BibitemOpen
  \bibfield  {author} {\bibinfo {author} {\bibfnamefont {M.}~\bibnamefont
  {Lappa}},\ }\href@noop {} {\emph {\bibinfo {title} {Thermal convection:
  patterns, evolution and stability}}}\ (\bibinfo  {publisher} {John Wiley \&
  Sons},\ \bibinfo {year} {2009})\BibitemShut {NoStop}%
\bibitem [{\citenamefont {Busse}(1978)}]{Busse1978}%
  \BibitemOpen
  \bibfield  {author} {\bibinfo {author} {\bibfnamefont {F.~H.}\ \bibnamefont
  {Busse}},\ }\bibfield  {title} {\bibinfo {title} {Non-linear properties of
  thermal convection},\ }\href {https://doi.org/10.1088/0034-4885/41/12/003}
  {\bibfield  {journal} {\bibinfo  {journal} {Reports on Progress in Physics}\
  }\textbf {\bibinfo {volume} {41}},\ \bibinfo {pages} {1929} (\bibinfo {year}
  {1978})}\BibitemShut {NoStop}%
\bibitem [{\citenamefont {Busse}\ and\ \citenamefont
  {Clever}(1979)}]{Busse1979}%
  \BibitemOpen
  \bibfield  {author} {\bibinfo {author} {\bibfnamefont {F.~H.}\ \bibnamefont
  {Busse}}\ and\ \bibinfo {author} {\bibfnamefont {R.~M.}\ \bibnamefont
  {Clever}},\ }\bibfield  {title} {\bibinfo {title} {Instabilities of
  convection rolls in a fluid of moderate prandtl number},\ }\href
  {https://doi.org/10.1017/S002211207900015X} {\bibfield  {journal} {\bibinfo
  {journal} {Journal of Fluid Mechanics}\ }\textbf {\bibinfo {volume} {91}},\
  \bibinfo {pages} {319–335} (\bibinfo {year} {1979})}\BibitemShut {NoStop}%
\bibitem [{\citenamefont {Subramanian}\ \emph {et~al.}(2016)\citenamefont
  {Subramanian}, \citenamefont {Brausch}, \citenamefont {Daniels},
  \citenamefont {Bodenschatz}, \citenamefont {Schneider},\ and\ \citenamefont
  {Pesch}}]{Subramanian2016}%
  \BibitemOpen
  \bibfield  {author} {\bibinfo {author} {\bibfnamefont {P.}~\bibnamefont
  {Subramanian}}, \bibinfo {author} {\bibfnamefont {O.}~\bibnamefont
  {Brausch}}, \bibinfo {author} {\bibfnamefont {K.~E.}\ \bibnamefont
  {Daniels}}, \bibinfo {author} {\bibfnamefont {E.}~\bibnamefont
  {Bodenschatz}}, \bibinfo {author} {\bibfnamefont {T.~M.}\ \bibnamefont
  {Schneider}},\ and\ \bibinfo {author} {\bibfnamefont {W.}~\bibnamefont
  {Pesch}},\ }\bibfield  {title} {\bibinfo {title} {Spatio-temporal patterns in
  inclined layer convection},\ }\href {https://doi.org/10.1017/jfm.2016.186}
  {\bibfield  {journal} {\bibinfo  {journal} {Journal of Fluid Mechanics}\
  }\textbf {\bibinfo {volume} {794}},\ \bibinfo {pages} {719} (\bibinfo {year}
  {2016})}\BibitemShut {NoStop}%
\bibitem [{\citenamefont {Reetz}\ and\ \citenamefont
  {Schneider}(2020)}]{Reetz2020e}%
  \BibitemOpen
  \bibfield  {author} {\bibinfo {author} {\bibfnamefont {F.}~\bibnamefont
  {Reetz}}\ and\ \bibinfo {author} {\bibfnamefont {T.~M.}\ \bibnamefont
  {Schneider}},\ }\bibfield  {title} {\bibinfo {title} {Invariant states in
  inclined layer convection. part 1. temporal transitions along dynamical
  connections between invariant states},\ }\bibfield  {journal} {\bibinfo
  {journal} {Journal of Fluid Mechanics}\ }\textbf {\bibinfo {volume} {898}},\
  \href {https://doi.org/10.1017/jfm.2020.317} {10.1017/jfm.2020.317} (\bibinfo
  {year} {2020})\BibitemShut {NoStop}%
\bibitem [{\citenamefont {Reetz}\ \emph {et~al.}(2020)\citenamefont {Reetz},
  \citenamefont {Subramanian},\ and\ \citenamefont {Schneider}}]{Reetz2020}%
  \BibitemOpen
  \bibfield  {author} {\bibinfo {author} {\bibfnamefont {F.}~\bibnamefont
  {Reetz}}, \bibinfo {author} {\bibfnamefont {P.}~\bibnamefont {Subramanian}},\
  and\ \bibinfo {author} {\bibfnamefont {T.~M.}\ \bibnamefont {Schneider}},\
  }\bibfield  {title} {\bibinfo {title} {Invariant states in inclined layer
  convection. part 2. bifurcations and connections between branches of
  invariant states},\ }\bibfield  {journal} {\bibinfo  {journal} {Journal of
  Fluid Mechanics}\ }\textbf {\bibinfo {volume} {898}},\ \href
  {https://doi.org/10.1017/jfm.2020.318} {10.1017/jfm.2020.318} (\bibinfo
  {year} {2020})\BibitemShut {NoStop}%
\bibitem [{\citenamefont {Zheng}\ \emph
  {et~al.}(2024{\natexlab{a}})\citenamefont {Zheng}, \citenamefont
  {Tuckerman},\ and\ \citenamefont {Schneider}}]{Zheng2024}%
  \BibitemOpen
  \bibfield  {author} {\bibinfo {author} {\bibfnamefont {Z.}~\bibnamefont
  {Zheng}}, \bibinfo {author} {\bibfnamefont {L.~S.}\ \bibnamefont
  {Tuckerman}},\ and\ \bibinfo {author} {\bibfnamefont {T.~M.}\ \bibnamefont
  {Schneider}},\ }\bibfield  {title} {\bibinfo {title} {Natural convection in a
  vertical channel. part 1. wavenumber interaction and eckhaus instability in a
  narrow domain},\ }\href {https://doi.org/10.1017/jfm.2024.842} {\bibfield
  {journal} {\bibinfo  {journal} {Journal of Fluid Mechanics}\ }\textbf
  {\bibinfo {volume} {1000}},\ \bibinfo {pages} {A28} (\bibinfo {year}
  {2024}{\natexlab{a}})}\BibitemShut {NoStop}%
\bibitem [{\citenamefont {Zheng}\ \emph
  {et~al.}(2024{\natexlab{b}})\citenamefont {Zheng}, \citenamefont
  {Tuckerman},\ and\ \citenamefont {Schneider}}]{Zheng2024b}%
  \BibitemOpen
  \bibfield  {author} {\bibinfo {author} {\bibfnamefont {Z.}~\bibnamefont
  {Zheng}}, \bibinfo {author} {\bibfnamefont {L.~S.}\ \bibnamefont
  {Tuckerman}},\ and\ \bibinfo {author} {\bibfnamefont {T.~M.}\ \bibnamefont
  {Schneider}},\ }\bibfield  {title} {\bibinfo {title} {Natural convection in a
  vertical channel. part 2. oblique solutions and global bifurcations in a
  spanwise-extended domain},\ }\href {https://doi.org/10.1017/jfm.2024.840}
  {\bibfield  {journal} {\bibinfo  {journal} {Journal of Fluid Mechanics}\
  }\textbf {\bibinfo {volume} {1000}},\ \bibinfo {pages} {A29} (\bibinfo {year}
  {2024}{\natexlab{b}})}\BibitemShut {NoStop}%
\bibitem [{\citenamefont {Qi}\ \emph {et~al.}(2019)\citenamefont {Qi},
  \citenamefont {Wang}, \citenamefont {Jaramillo-Villegas}, \citenamefont {Qi},
  \citenamefont {Weiner}, \citenamefont {D'Aguanno}, \citenamefont
  {Carruthers},\ and\ \citenamefont {Menyuk}}]{Qi2019}%
  \BibitemOpen
  \bibfield  {author} {\bibinfo {author} {\bibfnamefont {Z.}~\bibnamefont
  {Qi}}, \bibinfo {author} {\bibfnamefont {S.}~\bibnamefont {Wang}}, \bibinfo
  {author} {\bibfnamefont {J.}~\bibnamefont {Jaramillo-Villegas}}, \bibinfo
  {author} {\bibfnamefont {M.}~\bibnamefont {Qi}}, \bibinfo {author}
  {\bibfnamefont {A.~M.}\ \bibnamefont {Weiner}}, \bibinfo {author}
  {\bibfnamefont {G.}~\bibnamefont {D'Aguanno}}, \bibinfo {author}
  {\bibfnamefont {T.~F.}\ \bibnamefont {Carruthers}},\ and\ \bibinfo {author}
  {\bibfnamefont {C.~R.}\ \bibnamefont {Menyuk}},\ }\bibfield  {title}
  {\bibinfo {title} {{Dissipative cnoidal waves (Turing rolls) and the soliton
  limit in microring resonators}},\ }\href
  {https://doi.org/10.1364/OPTICA.6.001220} {\bibfield  {journal} {\bibinfo
  {journal} {Optica}\ }\textbf {\bibinfo {volume} {6}},\ \bibinfo {pages}
  {1220} (\bibinfo {year} {2019})},\ \Eprint {https://arxiv.org/abs/1905.07086}
  {arXiv:1905.07086} \BibitemShut {NoStop}%
\bibitem [{\citenamefont {Parra-Rivas}\ \emph {et~al.}(2018)\citenamefont
  {Parra-Rivas}, \citenamefont {Gomila}, \citenamefont {Gelens},\ and\
  \citenamefont {Knobloch}}]{Parra-Rivas2018a}%
  \BibitemOpen
  \bibfield  {author} {\bibinfo {author} {\bibfnamefont {P.}~\bibnamefont
  {Parra-Rivas}}, \bibinfo {author} {\bibfnamefont {D.}~\bibnamefont {Gomila}},
  \bibinfo {author} {\bibfnamefont {L.}~\bibnamefont {Gelens}},\ and\ \bibinfo
  {author} {\bibfnamefont {E.}~\bibnamefont {Knobloch}},\ }\bibfield  {title}
  {\bibinfo {title} {Bifurcation structure of periodic patterns in the
  lugiato-lefever equation with anomalous dispersion},\ }\href
  {https://doi.org/10.1103/PhysRevE.98.042212} {\bibfield  {journal} {\bibinfo
  {journal} {Physical Review E}\ }\textbf {\bibinfo {volume} {98}},\ \bibinfo
  {pages} {042212} (\bibinfo {year} {2018})}\BibitemShut {NoStop}%
\bibitem [{\citenamefont {{Parra-Rivas}}\ \emph {et~al.}(2018)\citenamefont
  {{Parra-Rivas}}, \citenamefont {Gomila}, \citenamefont {Gelens},\ and\
  \citenamefont {Knobloch}}]{parra-rivas2018BifurcationStructureLocalized}%
  \BibitemOpen
  \bibfield  {author} {\bibinfo {author} {\bibfnamefont {P.}~\bibnamefont
  {{Parra-Rivas}}}, \bibinfo {author} {\bibfnamefont {D.}~\bibnamefont
  {Gomila}}, \bibinfo {author} {\bibfnamefont {L.}~\bibnamefont {Gelens}},\
  and\ \bibinfo {author} {\bibfnamefont {E.}~\bibnamefont {Knobloch}},\
  }\bibfield  {title} {\bibinfo {title} {Bifurcation structure of localized
  states in the {{Lugiato-Lefever}} equation with anomalous dispersion},\
  }\href {https://doi.org/10.1103/PhysRevE.97.042204} {\bibfield  {journal}
  {\bibinfo  {journal} {Phys. Rev. E}\ }\textbf {\bibinfo {volume} {97}},\
  \bibinfo {pages} {042204} (\bibinfo {year} {2018})}\BibitemShut {NoStop}%
\bibitem [{\citenamefont {Barashenkov}\ \emph {et~al.}(1998)\citenamefont
  {Barashenkov}, \citenamefont {Smirnov},\ and\ \citenamefont
  {Alexeeva}}]{Barashenkov1998Bifurcation}%
  \BibitemOpen
  \bibfield  {author} {\bibinfo {author} {\bibfnamefont {I.~V.}\ \bibnamefont
  {Barashenkov}}, \bibinfo {author} {\bibfnamefont {Y.~S.}\ \bibnamefont
  {Smirnov}},\ and\ \bibinfo {author} {\bibfnamefont {N.~V.}\ \bibnamefont
  {Alexeeva}},\ }\bibfield  {title} {\bibinfo {title} {{Bifurcation to
  multisoliton complexes in the ac-driven, damped nonlinear Schr{\"{o}}dinger
  equation}},\ }\href {https://doi.org/10.1103/PhysRevE.57.2350} {\bibfield
  {journal} {\bibinfo  {journal} {Physical Review E - Statistical Physics,
  Plasmas, Fluids, and Related Interdisciplinary Topics}\ }\textbf {\bibinfo
  {volume} {57}},\ \bibinfo {pages} {2350} (\bibinfo {year}
  {1998})}\BibitemShut {NoStop}%
\bibitem [{\citenamefont {Godey}(2017)}]{Cyril2017}%
  \BibitemOpen
  \bibfield  {author} {\bibinfo {author} {\bibfnamefont {C.}~\bibnamefont
  {Godey}},\ }\bibfield  {title} {\bibinfo {title} {{A bifurcation analysis for
  the Lugiato-Lefever equation}},\ }\href
  {https://doi.org/10.1140/epjd/e2017-80057-2} {\bibfield  {journal} {\bibinfo
  {journal} {The European Physical Journal D}\ }\textbf {\bibinfo {volume}
  {71}},\ \bibinfo {pages} {131} (\bibinfo {year} {2017})}\BibitemShut
  {NoStop}%
\bibitem [{\citenamefont {Barashenkov}\ and\ \citenamefont
  {Smirnov}(1996)}]{Barashenkov1996Existence}%
  \BibitemOpen
  \bibfield  {author} {\bibinfo {author} {\bibfnamefont {I.~V.}\ \bibnamefont
  {Barashenkov}}\ and\ \bibinfo {author} {\bibfnamefont {Y.~S.}\ \bibnamefont
  {Smirnov}},\ }\bibfield  {title} {\bibinfo {title} {{Existence and stability
  chart for the ac-driven, damped nonlinear Schr{\"{o}}dinger solitons}},\
  }\href {https://doi.org/10.1103/PhysRevE.54.5707} {\bibfield  {journal}
  {\bibinfo  {journal} {Physical Review E}\ }\textbf {\bibinfo {volume} {54}},\
  \bibinfo {pages} {5707} (\bibinfo {year} {1996})}\BibitemShut {NoStop}%
\bibitem [{\citenamefont {Yelo-Sarri{\'{o}}n}\ \emph
  {et~al.}(2021)\citenamefont {Yelo-Sarri{\'{o}}n}, \citenamefont
  {Parra-Rivas}, \citenamefont {Englebert}, \citenamefont {Arab{\'{i}}},
  \citenamefont {Leo},\ and\ \citenamefont
  {Gorza}}]{Yelo-Sarrion2021Self-Pulsing}%
  \BibitemOpen
  \bibfield  {author} {\bibinfo {author} {\bibfnamefont {J.}~\bibnamefont
  {Yelo-Sarri{\'{o}}n}}, \bibinfo {author} {\bibfnamefont {P.}~\bibnamefont
  {Parra-Rivas}}, \bibinfo {author} {\bibfnamefont {N.}~\bibnamefont
  {Englebert}}, \bibinfo {author} {\bibfnamefont {C.~M.}\ \bibnamefont
  {Arab{\'{i}}}}, \bibinfo {author} {\bibfnamefont {F.}~\bibnamefont {Leo}},\
  and\ \bibinfo {author} {\bibfnamefont {S.-P.}\ \bibnamefont {Gorza}},\
  }\bibfield  {title} {\bibinfo {title} {{Self-pulsing in driven-dissipative
  photonic Bose-Hubbard dimers}},\ }\href
  {https://doi.org/10.1103/PhysRevResearch.3.L042031} {\bibfield  {journal}
  {\bibinfo  {journal} {Physical Review Research}\ }\textbf {\bibinfo {volume}
  {3}},\ \bibinfo {pages} {L042031} (\bibinfo {year} {2021})},\ \Eprint
  {https://arxiv.org/abs/2104.00649} {arXiv:2104.00649} \BibitemShut {NoStop}%
\bibitem [{\citenamefont {Yelo-Sarri{\'{o}}n}\ \emph
  {et~al.}(2022)\citenamefont {Yelo-Sarri{\'{o}}n}, \citenamefont {Leo},
  \citenamefont {Gorza},\ and\ \citenamefont
  {Parra-Rivas}}]{Yelo-Sarrion2022Self-Pulsing}%
  \BibitemOpen
  \bibfield  {author} {\bibinfo {author} {\bibfnamefont {J.}~\bibnamefont
  {Yelo-Sarri{\'{o}}n}}, \bibinfo {author} {\bibfnamefont {F.}~\bibnamefont
  {Leo}}, \bibinfo {author} {\bibfnamefont {S.-P.}\ \bibnamefont {Gorza}},\
  and\ \bibinfo {author} {\bibfnamefont {P.}~\bibnamefont {Parra-Rivas}},\
  }\bibfield  {title} {\bibinfo {title} {{Self-pulsing and chaos in the
  asymmetrically driven dissipative photonic Bose–Hubbard dimer: A
  bifurcation analysis}},\ }\href {https://doi.org/10.1063/5.0088597}
  {\bibfield  {journal} {\bibinfo  {journal} {Chaos: An Interdisciplinary
  Journal of Nonlinear Science}\ }\textbf {\bibinfo {volume} {32}},\ \bibinfo
  {pages} {1} (\bibinfo {year} {2022})},\ \Eprint
  {https://arxiv.org/abs/2202.04066} {arXiv:2202.04066} \BibitemShut {NoStop}%
\bibitem [{\citenamefont {Tusnin}\ \emph {et~al.}(2023)\citenamefont {Tusnin},
  \citenamefont {Tikan}, \citenamefont {Komagata},\ and\ \citenamefont
  {Kippenberg}}]{Tusnin2023Nonlinear}%
  \BibitemOpen
  \bibfield  {author} {\bibinfo {author} {\bibfnamefont {A.}~\bibnamefont
  {Tusnin}}, \bibinfo {author} {\bibfnamefont {A.}~\bibnamefont {Tikan}},
  \bibinfo {author} {\bibfnamefont {K.}~\bibnamefont {Komagata}},\ and\
  \bibinfo {author} {\bibfnamefont {T.~J.}\ \bibnamefont {Kippenberg}},\
  }\bibfield  {title} {\bibinfo {title} {{Nonlinear dynamics and Kerr frequency
  comb formation in lattices of coupled microresonators}},\ }\bibfield
  {journal} {\bibinfo  {journal} {Communications Physics}\ }\textbf {\bibinfo
  {volume} {6}},\ \href {https://doi.org/10.1038/s42005-023-01438-z}
  {10.1038/s42005-023-01438-z} (\bibinfo {year} {2023})\BibitemShut {NoStop}%
\bibitem [{\citenamefont {Tikan}\ \emph
  {et~al.}(2022{\natexlab{b}})\citenamefont {Tikan}, \citenamefont {Bonnefoy},
  \citenamefont {Ducrozet}, \citenamefont {Prabhudesai}, \citenamefont
  {Michel}, \citenamefont {Cazaubiel}, \citenamefont {Falcon}, \citenamefont
  {Copie}, \citenamefont {Randoux},\ and\ \citenamefont
  {Suret}}]{tikan2022NonlinearDispersionRelation}%
  \BibitemOpen
  \bibfield  {author} {\bibinfo {author} {\bibfnamefont {A.}~\bibnamefont
  {Tikan}}, \bibinfo {author} {\bibfnamefont {F.}~\bibnamefont {Bonnefoy}},
  \bibinfo {author} {\bibfnamefont {G.}~\bibnamefont {Ducrozet}}, \bibinfo
  {author} {\bibfnamefont {G.}~\bibnamefont {Prabhudesai}}, \bibinfo {author}
  {\bibfnamefont {G.}~\bibnamefont {Michel}}, \bibinfo {author} {\bibfnamefont
  {A.}~\bibnamefont {Cazaubiel}}, \bibinfo {author} {\bibfnamefont
  {{\'E}.}~\bibnamefont {Falcon}}, \bibinfo {author} {\bibfnamefont
  {F.}~\bibnamefont {Copie}}, \bibinfo {author} {\bibfnamefont
  {S.}~\bibnamefont {Randoux}},\ and\ \bibinfo {author} {\bibfnamefont
  {P.}~\bibnamefont {Suret}},\ }\bibfield  {title} {\bibinfo {title} {Nonlinear
  dispersion relation in integrable turbulence},\ }\href
  {https://doi.org/10.1038/s41598-022-14209-7} {\bibfield  {journal} {\bibinfo
  {journal} {Sci Rep}\ }\textbf {\bibinfo {volume} {12}},\ \bibinfo {pages}
  {10386} (\bibinfo {year} {2022}{\natexlab{b}})}\BibitemShut {NoStop}%
\bibitem [{\citenamefont {Leisman}\ \emph {et~al.}(2019)\citenamefont
  {Leisman}, \citenamefont {Zhou}, \citenamefont {Banks}, \citenamefont
  {Kova{\v c}i{\v c}},\ and\ \citenamefont
  {Cai}}]{leisman2019EffectiveDispersionFocusing}%
  \BibitemOpen
  \bibfield  {author} {\bibinfo {author} {\bibfnamefont {K.~P.}\ \bibnamefont
  {Leisman}}, \bibinfo {author} {\bibfnamefont {D.}~\bibnamefont {Zhou}},
  \bibinfo {author} {\bibfnamefont {J.~W.}\ \bibnamefont {Banks}}, \bibinfo
  {author} {\bibfnamefont {G.}~\bibnamefont {Kova{\v c}i{\v c}}},\ and\
  \bibinfo {author} {\bibfnamefont {D.}~\bibnamefont {Cai}},\ }\bibfield
  {title} {\bibinfo {title} {Effective dispersion in the focusing nonlinear
  {{Schr}}\"odinger equation},\ }\href
  {https://doi.org/10.1103/PhysRevE.100.022215} {\bibfield  {journal} {\bibinfo
   {journal} {Phys. Rev. E}\ }\textbf {\bibinfo {volume} {100}},\ \bibinfo
  {pages} {022215} (\bibinfo {year} {2019})}\BibitemShut {NoStop}%
\bibitem [{\citenamefont {Lee}\ \emph {et~al.}(2009{\natexlab{a}})\citenamefont
  {Lee}, \citenamefont {Kova{\v{c}}i{\v{c}}},\ and\ \citenamefont
  {Cai}}]{Lee2009Renormalized}%
  \BibitemOpen
  \bibfield  {author} {\bibinfo {author} {\bibfnamefont {W.}~\bibnamefont
  {Lee}}, \bibinfo {author} {\bibfnamefont {G.}~\bibnamefont
  {Kova{\v{c}}i{\v{c}}}},\ and\ \bibinfo {author} {\bibfnamefont
  {D.}~\bibnamefont {Cai}},\ }\bibfield  {title} {\bibinfo {title}
  {{Renormalized Resonance Quartets in Dispersive Wave Turbulence}},\ }\href
  {https://doi.org/10.1103/PhysRevLett.103.024502} {\bibfield  {journal}
  {\bibinfo  {journal} {Physical Review Letters}\ }\textbf {\bibinfo {volume}
  {103}},\ \bibinfo {pages} {2} (\bibinfo {year}
  {2009}{\natexlab{a}})}\BibitemShut {NoStop}%
\bibitem [{\citenamefont {Komagata}\ \emph {et~al.}(2021)\citenamefont
  {Komagata}, \citenamefont {Tikan}, \citenamefont {Tusnin}, \citenamefont
  {Riemensberger}, \citenamefont {Churaev}, \citenamefont {Guo},\ and\
  \citenamefont {Kippenberg}}]{komagata2021dissipative}%
  \BibitemOpen
  \bibfield  {author} {\bibinfo {author} {\bibfnamefont {K.}~\bibnamefont
  {Komagata}}, \bibinfo {author} {\bibfnamefont {A.}~\bibnamefont {Tikan}},
  \bibinfo {author} {\bibfnamefont {A.}~\bibnamefont {Tusnin}}, \bibinfo
  {author} {\bibfnamefont {J.}~\bibnamefont {Riemensberger}}, \bibinfo {author}
  {\bibfnamefont {M.}~\bibnamefont {Churaev}}, \bibinfo {author} {\bibfnamefont
  {H.}~\bibnamefont {Guo}},\ and\ \bibinfo {author} {\bibfnamefont {T.~J.}\
  \bibnamefont {Kippenberg}},\ }\bibfield  {title} {\bibinfo {title}
  {Dissipative kerr solitons in a photonic dimer on both sides of exceptional
  point},\ }\bibfield  {journal} {\bibinfo  {journal} {Communications Physics}\
  }\href {https://doi.org/10.1038/s42005-021-00661-w}
  {10.1038/s42005-021-00661-w} (\bibinfo {year} {2021})\BibitemShut {NoStop}%
\bibitem [{\citenamefont {Lee}\ \emph {et~al.}(2009{\natexlab{b}})\citenamefont
  {Lee}, \citenamefont {Kova\ifmmode \check{c}\else
  \v{c}\fi{}i\ifmmode~\check{c}\else \v{c}\fi{}},\ and\ \citenamefont
  {Cai}}]{Lee2009}%
  \BibitemOpen
  \bibfield  {author} {\bibinfo {author} {\bibfnamefont {W.}~\bibnamefont
  {Lee}}, \bibinfo {author} {\bibfnamefont {G.}~\bibnamefont {Kova\ifmmode
  \check{c}\else \v{c}\fi{}i\ifmmode~\check{c}\else \v{c}\fi{}}},\ and\
  \bibinfo {author} {\bibfnamefont {D.}~\bibnamefont {Cai}},\ }\bibfield
  {title} {\bibinfo {title} {Renormalized resonance quartets in dispersive wave
  turbulence},\ }\href {https://doi.org/10.1103/PhysRevLett.103.024502}
  {\bibfield  {journal} {\bibinfo  {journal} {Phys. Rev. Lett.}\ }\textbf
  {\bibinfo {volume} {103}},\ \bibinfo {pages} {024502} (\bibinfo {year}
  {2009}{\natexlab{b}})}\BibitemShut {NoStop}%
\bibitem [{\citenamefont {Anderson}\ \emph {et~al.}(2023)\citenamefont
  {Anderson}, \citenamefont {Tikan}, \citenamefont {Tusnin}, \citenamefont
  {Riemensberger}, \citenamefont {Davydova}, \citenamefont {Wang},\ and\
  \citenamefont {Kippenberg}}]{Anderson2023}%
  \BibitemOpen
  \bibfield  {author} {\bibinfo {author} {\bibfnamefont {M.~H.}\ \bibnamefont
  {Anderson}}, \bibinfo {author} {\bibfnamefont {A.}~\bibnamefont {Tikan}},
  \bibinfo {author} {\bibfnamefont {A.}~\bibnamefont {Tusnin}}, \bibinfo
  {author} {\bibfnamefont {J.}~\bibnamefont {Riemensberger}}, \bibinfo {author}
  {\bibfnamefont {A.}~\bibnamefont {Davydova}}, \bibinfo {author}
  {\bibfnamefont {R.~N.}\ \bibnamefont {Wang}},\ and\ \bibinfo {author}
  {\bibfnamefont {T.~J.}\ \bibnamefont {Kippenberg}},\ }\bibfield  {title}
  {\bibinfo {title} {{Dissipative Solitons and Switching Waves in
  Dispersion-Modulated Kerr Cavities}},\ }\href
  {https://doi.org/10.1103/physrevx.13.011040} {\bibfield  {journal} {\bibinfo
  {journal} {Physical Review X}\ }\textbf {\bibinfo {volume} {13}},\ \bibinfo
  {pages} {11040} (\bibinfo {year} {2023})},\ \Eprint
  {https://arxiv.org/abs/2205.09957} {arXiv:2205.09957} \BibitemShut {NoStop}%
\bibitem [{\citenamefont {Burns}\ \emph {et~al.}(2020)\citenamefont {Burns},
  \citenamefont {Vasil}, \citenamefont {Oishi}, \citenamefont {Lecoanet},\ and\
  \citenamefont {Brown}}]{Burns2020}%
  \BibitemOpen
  \bibfield  {author} {\bibinfo {author} {\bibfnamefont {K.~J.}\ \bibnamefont
  {Burns}}, \bibinfo {author} {\bibfnamefont {G.~M.}\ \bibnamefont {Vasil}},
  \bibinfo {author} {\bibfnamefont {J.~S.}\ \bibnamefont {Oishi}}, \bibinfo
  {author} {\bibfnamefont {D.}~\bibnamefont {Lecoanet}},\ and\ \bibinfo
  {author} {\bibfnamefont {B.~P.}\ \bibnamefont {Brown}},\ }\bibfield  {title}
  {\bibinfo {title} {Dedalus: A flexible framework for numerical simulations
  with spectral methods},\ }\href
  {https://doi.org/10.1103/PhysRevResearch.2.023068} {\bibfield  {journal}
  {\bibinfo  {journal} {Physical Review Research}\ }\textbf {\bibinfo {volume}
  {2}},\ \bibinfo {pages} {023068} (\bibinfo {year} {2020})}\BibitemShut
  {NoStop}%
\bibitem [{\citenamefont {Gibson}(2012)}]{channelflow}%
  \BibitemOpen
  \bibfield  {author} {\bibinfo {author} {\bibfnamefont {J.~F.}\ \bibnamefont
  {Gibson}},\ }\href {Channelflow.org} {\bibinfo {title} {Channelflow: A
  spectral navier-stokes simulator in c++}} (\bibinfo {year}
  {2012})\BibitemShut {NoStop}%
\bibitem [{\citenamefont {Saad}\ and\ \citenamefont
  {Schultz}(1986)}]{Saad1986}%
  \BibitemOpen
  \bibfield  {author} {\bibinfo {author} {\bibfnamefont {Y.}~\bibnamefont
  {Saad}}\ and\ \bibinfo {author} {\bibfnamefont {M.~H.}\ \bibnamefont
  {Schultz}},\ }\bibfield  {title} {\bibinfo {title} {{GMRES: A Generalized
  Minimal Residual Algorithm for Solving Nonsymmetric Linear Systems}},\ }\href
  {https://doi.org/10.1137/0907058} {\bibfield  {journal} {\bibinfo  {journal}
  {SIAM Journal on Scientific and Statistical Computing}\ }\textbf {\bibinfo
  {volume} {7}},\ \bibinfo {pages} {856} (\bibinfo {year} {1986})}\BibitemShut
  {NoStop}%
\bibitem [{\citenamefont {Viswanath}(2007)}]{Viswanath2007}%
  \BibitemOpen
  \bibfield  {author} {\bibinfo {author} {\bibfnamefont {D.}~\bibnamefont
  {Viswanath}},\ }\bibfield  {title} {\bibinfo {title} {Recurrent motions
  within plane couette turbulence},\ }\href
  {https://doi.org/10.1017/S0022112007005459} {\bibfield  {journal} {\bibinfo
  {journal} {Journal of Fluid Mechanics}\ }\textbf {\bibinfo {volume} {580}},\
  \bibinfo {pages} {339} (\bibinfo {year} {2007})}\BibitemShut {NoStop}%
\bibitem [{\citenamefont {Parra-Rivas}\ \emph {et~al.}(2016)\citenamefont
  {Parra-Rivas}, \citenamefont {Knobloch}, \citenamefont {Gomila},\ and\
  \citenamefont {Gelens}}]{Parra-Rivas2016}%
  \BibitemOpen
  \bibfield  {author} {\bibinfo {author} {\bibfnamefont {P.}~\bibnamefont
  {Parra-Rivas}}, \bibinfo {author} {\bibfnamefont {E.}~\bibnamefont
  {Knobloch}}, \bibinfo {author} {\bibfnamefont {D.}~\bibnamefont {Gomila}},\
  and\ \bibinfo {author} {\bibfnamefont {L.}~\bibnamefont {Gelens}},\
  }\bibfield  {title} {\bibinfo {title} {{Dark solitons in the Lugiato-Lefever
  equation with normal dispersion}},\ }\href
  {https://doi.org/10.1103/PhysRevA.93.063839} {\bibfield  {journal} {\bibinfo
  {journal} {Physical Review A}\ }\textbf {\bibinfo {volume} {93}},\ \bibinfo
  {pages} {1} (\bibinfo {year} {2016})},\ \Eprint
  {https://arxiv.org/abs/1603.03985} {arXiv:1603.03985} \BibitemShut {NoStop}%
\bibitem [{\citenamefont {Burke}\ and\ \citenamefont
  {Knobloch}(2006)}]{Burke2006}%
  \BibitemOpen
  \bibfield  {author} {\bibinfo {author} {\bibfnamefont {J.}~\bibnamefont
  {Burke}}\ and\ \bibinfo {author} {\bibfnamefont {E.}~\bibnamefont
  {Knobloch}},\ }\bibfield  {title} {\bibinfo {title} {{Localized states in the
  generalized Swift-Hohenberg equation}},\ }\bibfield  {journal} {\bibinfo
  {journal} {Physical Review E - Statistical, Nonlinear, and Soft Matter
  Physics}\ }\textbf {\bibinfo {volume} {73}},\ \href
  {https://doi.org/10.1103/PhysRevE.73.056211} {10.1103/PhysRevE.73.056211}
  (\bibinfo {year} {2006})\BibitemShut {NoStop}%
\bibitem [{\citenamefont {Parra-Rivas}\ \emph {et~al.}(2014)\citenamefont
  {Parra-Rivas}, \citenamefont {Gomila}, \citenamefont {Mat{\'{i}}as},
  \citenamefont {Coen},\ and\ \citenamefont {Gelens}}]{Parra-Rivas2014}%
  \BibitemOpen
  \bibfield  {author} {\bibinfo {author} {\bibfnamefont {P.}~\bibnamefont
  {Parra-Rivas}}, \bibinfo {author} {\bibfnamefont {D.}~\bibnamefont {Gomila}},
  \bibinfo {author} {\bibfnamefont {M.~A.}\ \bibnamefont {Mat{\'{i}}as}},
  \bibinfo {author} {\bibfnamefont {S.}~\bibnamefont {Coen}},\ and\ \bibinfo
  {author} {\bibfnamefont {L.}~\bibnamefont {Gelens}},\ }\bibfield  {title}
  {\bibinfo {title} {{Dynamics of localized and patterned structures in the
  Lugiato-Lefever equation determine the stability and shape of optical
  frequency combs}},\ }\href {https://doi.org/10.1103/PhysRevA.89.043813}
  {\bibfield  {journal} {\bibinfo  {journal} {Physical Review A - Atomic,
  Molecular, and Optical Physics}\ }\textbf {\bibinfo {volume} {89}},\ \bibinfo
  {pages} {1} (\bibinfo {year} {2014})},\ \Eprint
  {https://arxiv.org/abs/1401.6059} {arXiv:1401.6059} \BibitemShut {NoStop}%
\end{thebibliography}
\end{document}